\documentclass[nofootinbib,amsmath,amssymb,10pt,eqsecnum, onecolumn]{revtex4}
\usepackage{graphicx, amsmath, amsmath, amssymb}
\usepackage[caption=false]{subfig}
\usepackage[breaklinks, colorlinks, citecolor=blue]{hyperref}
\usepackage{color}
\usepackage{natbib,hyperref,ifthen}

\begin{document}
\title{Tension between the power spectrum of density perturbations  measured on large and small scales}
\author{Richard A. Battye}
\email{Richard.Battye@manchester.ac.uk}
\affiliation{Jodrell Bank Centre for Astrophysics, School of Physics and Astronomy, University of Manchester, Manchester, M13 9PL, U.K.}
\author{Tom Charnock}
\email{Tom.Charnock@nottingham.ac.uk}
\affiliation{Centre for Astronomy \& Particle Theory, University of Nottingham, University Park, Nottingham, NG7 2RD, U.K.}

\author{Adam Moss}
\email{Adam.Moss@nottingham.ac.uk}
\affiliation{Centre for  Astronomy \& Particle Theory, University of Nottingham, University Park, Nottingham, NG7 2RD, U.K.}
\vskip 0.5cm
%\date{\today}

\begin{abstract}
There is a tension between measurements of the amplitude of the  power spectrum of density perturbations inferred using the Cosmic Microwave Background (CMB) and directly measured by Large-Scale Structure (LSS)  on smaller scales. We show that this tension exists, and is robust, for a range of LSS indicators including clusters, lensing and redshift space distortions and using CMB data from either {\it Planck} or WMAP+SPT/ACT. One obvious way to try to reconcile this is the inclusion of a massive neutrino which could be either active or sterile. Using {\it Planck} and a combination of all the LSS data we find that (i) for an active neutrino $\sum m_{\nu}= (0.357\pm0.099)\,{\rm eV}$ and (ii) for a sterile neutrino $m_{\rm sterile}^{\rm eff}= (0.67\pm0.18)\,{\rm eV}$ and $\Delta N_{\rm eff}= 0.32\pm0.20$. This is, however, at the expense of a degraded fit to {\em Planck} temperature data, and we quantify the residual tension at $2.5  \sigma$ and $1.6 \sigma$  for massive and sterile neutrinos respectively.  We also consider alternative explanations including a lower redshift for reionization that would be in conflict with polarisation measurements made by WMAP and {\it ad-hoc} modifications to primordial power spectrum. 
\end{abstract}

\maketitle

\section{Introduction} \label{sec:introduction}

A standard cosmological model has become established over the past few decades known as the $\Lambda$CDM model. It is specified by 6 cosmological parameters which are: the angular diameter of the Cosmic Microwave Background (CMB) acoustic scale, $\Theta_{\rm MC}$; the physical densities of cold dark matter and baryons measured relative to the critical density, $\Omega_{\rm c}h^2$ and $\Omega_{\rm b}h^2$ respectively; the amplitude, $A_{\rm S}$, and spectral index, $n_{\rm S}$, of primordial density perturbations; and the optical depth to reionizaton, $\tau_{\rm R}$. These can be converted into more conventional quantities such as the Hubble constant, $H_0=100\,h\,{\rm km}\,{\rm sec}^{-1}\,{\rm Mpc}^{-1}$, the densities of matter and the cosmological constant, $\Lambda$, relative to critical, $\Omega_{\rm m}$ and $\Omega_{\Lambda}$, and the redshift of reionization, $z_{\rm re}$. The $\Lambda$CDM model appears to provide an excellent fit to the wide range of data that has been gathered over the last few years. The race is on to accumulate evidence for extensions to this model and a contribution to this goal is the objective of this paper.

Our approach will be to compare measurements of the spectrum of anisotropies of the Cosmic Microwave Background (CMB), that typically probe large scales, with those from Large-Scale Structure (LSS),  that probe smaller scales\footnote[1]{We note that present CMB and LSS measurements do overlap, but the dichotomy between large and small scales that we suggest here is still fundamentally correct.}. Such an approach has a venerable history in the development of the $\Lambda$CDM model. Although the clinching piece of evidence was the detection of cosmic acceleration using type IA supernovae~\cite{1998AJ....116.1009R,1999ApJ...517..565P}, there had been clear indications that a Universe with critical matter density did not fit the data. One simple way of seeing this was that the value of $\sigma_8$, the r.m.s. perturbation in spheres of radius $8h^{-1}\,{\rm Mpc}$, for the COBE normalised critical matter density models was $\approx 1.5$, whereas observations from a range of indicators suggested that it was in the range $0.7-0.9$ favouring $\Omega_{\rm m}<1$~\cite{1997ApJ...489L...1H,1998MNRAS.298.1145E}. The shape of the matter power spectrum was also in conflict with $\Omega_{\rm m}=1$~\cite{1990Natur.348..705E,1999MNRAS.310..565B}.

In this paper we will perform a detailed study of recent published observations that we believe challenge the standard cosmological model and, under the strong presumption that all the data used is accurate and free from systematics, suggest that the presently established concordance model needs to be supplemented by new physics. The basic approach is similar to that described above: we will compare the best-fit cosmological models predicted by CMB measurements from the {\it Planck} satellite~\cite{2013arXiv1303.5062P}, the WiIkinson Microwave Anisotropy Probe (WMAP)~\cite{2013ApJS..208...20B}, the Atacama Cosmology Telescope (ACT)~\cite{Das:2013zf} and the South Pole Telescope (SPT)~\cite{2012ApJ...755...70R} with those predicted by probes of LSS. We will find a discrepancy between them that can be easily understood by plotting the best-fit contours in the $\sigma_8-\Omega_{\rm m}$ plane. This discrepancy is much smaller than that between the $\Lambda$CDM model and one with a critical matter density as described above. Nonetheless, we will show that it statistically significant and robust to different combinations of CMB and LSS data.

This paper follows up our recent work on massive neutrinos presented in \cite{Battye:2013} (see also \cite{Wyman:2013,Hamann:2013} for other combinations of data leading to a similar result). There it was shown that both SZ cluster counts and lensing, from both the CMB and cosmic shear, were in conflict with CMB measurements and that a neutrino component -- which could be from active or sterile neutrinos -- could be used to reconcile these measurements. In turn this built on the earlier suggestion in \cite{2013arXiv1303.5080P} that tension between the CMB measurements and SZ cluster counts could be accounted for in this way. The fact that the SZ cluster counts and the lensing data are compatible with each other strengthens the two $\sim 2\sigma$ discrepancies into a statistically improbable discrepancy of $\sim 4\sigma$. This reconciliation of measurements of large and small scales was at the expense of less good fit to the CMB data -- the two being seen to be in conflict at the level of $2.8\sigma$~\cite{Battye:2013}. 

Neutrino masses are an obvious way to explain a dearth of power on small scales. Particle physics oscillation experiments are sensitive to the square differences between the neutrino masses and cosmology is mostly sensitive to the sum of the masses, $\sum m_{\nu}$. Using our results, the required values would suggest either an inverted/degenerate hierarchy or  an extra neutrino that would need to be sterile.  If the neutrinos are sterile it is possible that the best-fit model can be made more compatible with direct measurements of the Hubble constant from low-redshift standard candles such as Cepheids~\cite{Riess:2011yx}. 

The preference for massive neutrinos reported in \cite{Battye:2013, Wyman:2013,Hamann:2013} is a result of a global fit with an extended cosmological model. Clearly any systematic error in the data, or its interpretation, could lead to a false detection and therefore it needs to be treated with caution. Moreover, there are other extensions to the standard model that could lead to a similar result. We will explore a range of different data combinations and also a non-exhaustive range of alternative explanations for the tension, in particular optical depth assumptions inferred from WMAP polarization, and modifications to the primordial power spectrum.  We find that although there is strong evidence for a discrepancy, in excess of $5\sigma$, no one model is able to improve to fit to each individual likelihood component.

The data we will use in this paper are listed below.

\begin{itemize}
\item {\bf  {\it Planck} + WP}: We use measurements of the temperature anisotropy power spectrum made by {\it Planck}  that have been extensively used for cosmological parameter analysis~\cite{2013arXiv1303.5076P}.  These results cover the multipole range $\ell\approx 2-2500$. This is implemented by using the standard likelihood~\cite{2013arXiv1303.5075P} and uses the measurements of the polarisation and temperature-polarisation cross-correlation power spectra from WMAP 9 year data~\cite{2013ApJS..208...19H}.

\item{\bf WMAP + highL}: As an alternative to the {\it Planck} temperature measurements, we will use those made by WMAP~\cite{2013ApJS..208...19H}  over the multipole range $\ell\approx 2-800$ complemented with  higher resolution measurements made by ACT~\cite{Das:2013zf} and SPT~\cite{2012ApJ...755...70R}. As we will see in the subsequent discussion there are subtle quantitative  differences between the conclusions that one draws if one makes this choice, but the qualitative results are the same.

\item{\bf Baryonic Acoustic Oscillations (BAO)}: The ratio of the sound horizon at the drag epoch, $r_{\rm s}(z_{\rm d})$, to the volume-averaged distance $D_{\rm V}(z_{\rm eff})$, can be constrained using BAO. We use the results of several surveys which detect the BAO signal in the power spectrum: (1) the 6dF Galaxy survey~\cite{Jones:2004zy,Beutler:2011hx}, which constrains $D_{\rm V}(z_{\rm eff})=(456\pm27)\,{\rm Mpc}$ and $r_{\rm s}(z_{\rm d})/D_{\rm V}(z_{\rm eff})=0.336\pm0.015$ (4.5\% precision) where $z_{\rm eff}=0.106$; (2) the SDSS DR7 measurement at  $z_{\rm eff}=0.35$ as reanalysed by~\cite{Padmanabhan:2012hf}, which constrains $D_{\rm V}(z_{\rm eff})/r_{\rm s}(z_{\rm d})=8.88\pm0.17$; (3) the Baryon Oscillation Spectroscopic Survey (BOSS), which maps the spatial distribution of luminous red galaxies and quasars to detect the characteristic BAO scale. The Data Release 9 (DR9) results constrain $D_{\rm V}(z_{\rm eff})/r_{\rm s}(z_{\rm d})=13.67\pm0.22$ at $z_{\rm eff}=0.57$~\cite{Anderson:2012sa}.

\item{\bf SZ Cluster counts}: {\it Planck} has detected clusters via the Sunyaev-Zeldovich effect. Using a sample of 189 clusters, cosmological constraints were deduced in the $\sigma_8-\Omega_{\rm m}$ plane~\cite{2013arXiv1303.5080P}. We will implement this using two priors: $\sigma_8(\Omega_{\rm m}/0.27)^{0.3}=0.78\pm 0.01$ which corresponds to a fixed hydrostatic bias of $1-b=0.8$, and $\sigma_8(\Omega_{\rm m}/0.27)^{0.3}=0.764 \pm 0.025$, where $1-b$ is allowed to vary in the range $[0.7,1.0]$. This is compatible with other determinations using cluster counts selected using the SZ effect~\cite{Saro:2013,Hasselfield:2013} and in other wavebands~\cite{vik09,2010ApJ...708..645R}.

\item{\bf CFHTLens and CMB lensing from {\it Planck} and SPT}: We will use two types of weak lensing measurements that we will denote as ``lensing" in the subsequent discussion. The first is the coherent distortion of galaxy shapes, sometimes called cosmic shear, measured by the Canada France Hawaii Telescope Lensing Survey (CFHTLenS)~\cite{2012MNRAS.427..146H}. In this analysis we have used the tomographic blue galaxy sample, which was shown in \cite{Heymans:2013fya} to have an intrinsic alignment signal that was consistent with zero.  This eliminates the need to marginalise over any additional nuisance parameters.  The cosmic shear correlation functions are estimated in six redshift bins, each with an angular range $1.5 < \theta < 35$ arcmin. We have tested this dataset gives consistent results to the 2D data used in our previous work~\cite{2013MNRAS.430.2200K}. As in that analysis, we correct the power spectrum on non-linear scales using the {\tt Halofit} fitting formulae~\cite{Smith:2002dz,Takahashi:2012em}, which has been shown to be accurate enough to use with massive neutrinos~\cite{Bird:2011rb}. In addition to cosmic shear we use measurements of the lensing of the CMB. This has been detected over a wide range of scales and we use reconstructions from {\it Planck}~\cite{2013arXiv1303.5077P} and SPT~\cite{2012ApJ...756..142V}.

\item{\bf Redshift space distortions from BOSS (RSD)}: When non-linear effects are taken into account it is possible to garner information on RSD and the Alcock-Paczynski (AP) effect from the anisotropic galaxy power spectrum in redshift space. Modelling this involves introducing distortion parameters to account for the anisotropy caused by the deviation from the fiducial cosmological parameters used to convert galaxy redshifts from that of the real cosmology. The Alcock-Paczynski effect is sensitive to 
\begin{equation}
  F_{\rm AP}(z_{\rm eff})=\frac{\alpha_\parallel}{\alpha_\perp}(1+z)\frac{D_{\rm A}^{\rm fid}(z)H^{\rm fid}(z)}{c}\,,
 \end{equation}
  whilst the BAO signal measures 
  \begin{equation}
  \frac{D_{\rm V}(z_{\rm eff})}{r_{\rm s}(z_{\rm d})}=\frac{\left(\alpha_\perp^2\alpha_\parallel[(1+z_{\rm eff})D_{\rm A}^{\rm fid}(z_{\rm eff})]^2\dfrac{cz_{\rm eff}}{H^{\rm fid}(z_{\rm eff})}\right)^{1/3}}{r_{\rm s}^{\rm fid}(z_{\rm d})}\,,
  \end{equation}
  where $\alpha_\parallel$ and $\alpha_\perp$ are the scaling factors  along the line-of-sight direction and perpendicular to it respectively. $D_{\rm A}^{\rm eff}(z)$, $H^{\rm fid}(z)$ and $r_{\rm s}^{\rm fid}(z_{\rm d})$ are fiducial values of the angular diameter distance, the Hubble parameter and the distance to the sound horizon during the drag epoch.
  Using $F_{\rm AP}(z_{\rm eff})$ and $D_{\rm V}(z_{\rm eff})/r_{\rm s}(z_{\rm d})$ the degeneracy between $D_{\rm A}(z)$ and $H(z)$ is broken. Finally, the relative amplitude between the monopole and the quadrupole constrains the growth rate $f(z)\sigma_8(z)$. For BOSS DR11 RSD measurements, these parameters and their covariance estimated with $k_{\rm max}=0.20h\,{\rm Mpc}^{-1}$ are~\cite{Beutler:2013yhm}
  \begin{equation}
  \begin{pmatrix}D_{\rm V}(z_{\rm eff})/r_{\rm s}(z_{\rm d})\\F_{\rm AP}(z_{\rm eff})\\f(z_{\rm eff})\sigma_8(z_{\rm eff})\end{pmatrix}=\begin{pmatrix}13.88\\0.683\\0.422\end{pmatrix}\phantom{ and }C_{k_{\rm eff}=0.20}^{-1}=\begin{pmatrix}31.032&77.773&-16.796\\&2687.7&-1475.9\\&&\phantom{-}1323.0\end{pmatrix}.
  \end{equation}
  We use these intermediate data products in our Markov Chain Monte Carlo  (MCMC) analysis. The RSD results have also been shown to be accurate enough to use with models with non-zero neutrino mass~\cite{Beutler:2014yhv}.

\end{itemize}

Implementation of the RSDs and the use of the tomographic lensing data from CFHTLenS are the improvements over the previous analysis presented in \cite{Battye:2013}. We also consider the more conservative {\it Planck} SZ result obtained with  $1-b=[0.7,1.0]$, where previously this was fixed at $1-b=0.8$ in~\cite{Battye:2013}. We note that the impact of RSD data on constraints on massive neutrinos was considered in \cite{Beutler:2014yhv}. There the CFHTLenS data was implemented via a prior in the $\sigma_8-\Omega_{\rm m}$ plane and the impact of clusters was not considered. Qualitatively our results are compatible with those presented in that paper.

The remainder of the paper is organised as follows. In section \ref{sec:tension} we will make the case that there is a tension between the CMB and LSS measurements and then in \ref{sec:neutrinos} we perform an extensive study of neutrinos as its resolution. In section \ref{sec:alternative} we consider a number of alternative explanations before providing discussion in section~\ref{sec:discussion} and concluding in \ref{sec:conclusions}.

\section{Case for a tension between measurements of the Cosmic Microwave Background and Large-Scale Structure}
\label{sec:tension}

\begin{figure}
    \centering
    \mbox{\resizebox{0.6\textwidth}{!}{\includegraphics{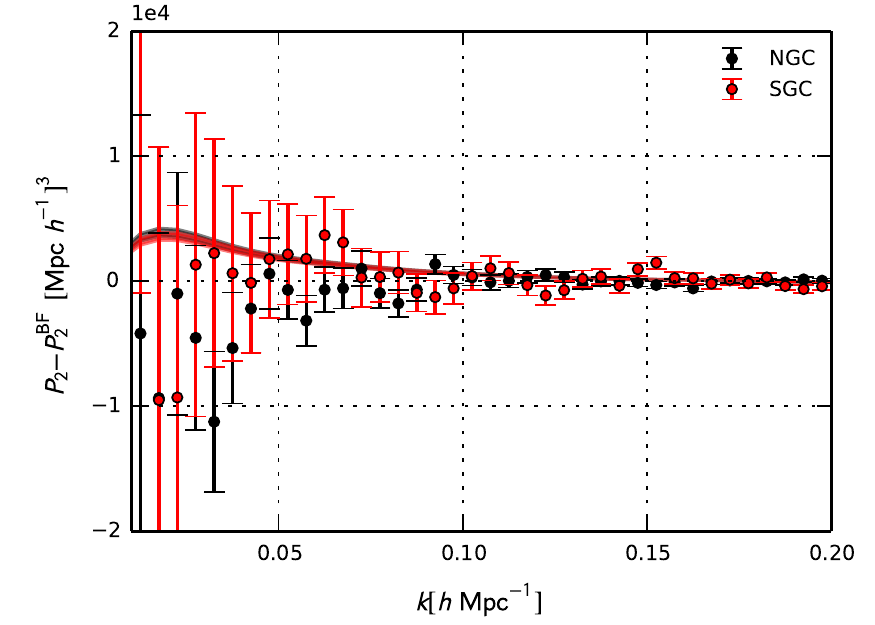}}}
    \caption{The quadrupole component of the redshift space power spectrum for the range of allowed {\it Planck} cosmologies (indicated by the narrow red and black bands), relative to the best-fit BOSS spectrum, for the North and South Galactic Caps (NGC and SGC respectively).}
    \label{fig:RSDcomp}
\end{figure}

In this section we will make the case that there is a discrepancy between CMB measurements and those from LSS. Note that this is a {\em model independent} statement within the context of $\Lambda$CDM. We will do this by first considering each of the {\em individual} LSS data. The observed SZ cluster counts are a factor $\sim 2$ lower than would be expected in the cosmological models preferred by CMB data~\cite{2013arXiv1303.5076P} and it was shown in \cite{Battye:2013} that both the cosmic shear and CMB lensing data are systematically low compared to the same expectations. In Fig.~\ref{fig:RSDcomp} we present a similar visual illustration of the discrepancy between the allowed {\it Planck} cosmologies and the RSD data. Here the theoretical RSD multipole power spectra were computed according to the same procedure as in~\cite{Beutler:2013yhm}.  First, corrections to the linear {\tt CAMB}~\cite{2000ApJ...538..473L} power spectrum were applied at 2-loop order using the {\tt RegPT} code~\cite{Taruya:2012ut}.  The anisotropic galaxy power spectrum was then modelled using the prescription in~\cite{Taruya:2010mx}, which includes corrections due to the coupling between the density and velocity components. Next, we apply bias corrections to the density field according to \cite{McDonald:2009dh}, and finally, apply window functions for the North and South Galactic Caps (NGC and SGC respectively). In Fig.~\ref{fig:RSDcomp} we show the quadrupole component of the anisotropic power spectrum for {\it Planck} $\Lambda$CDM cosmologies, relative to the best-fit BOSS spectra (where the growth rate is treated as a free parameter) for the NGC and SGC. For the purposes of the plot we have fixed the bias parameters (which are nuisance parameters in the full fit) to their BOSS best-fit values, but included the range of {\it Planck} cosmologies allowed by an MCMC analysis.  The excess power (on large scales) is again apparent for the {\it Planck} cosmologies, visually showing the preference of the BOSS data for a lower growth rate.  In practice calculating the non-linear corrections are computationally expensive, so we use the intermediate BOSS data and covariance matrix products given in section~\ref{sec:introduction} for MCMC fitting. 

\begin{figure}
  \centering
  \includegraphics{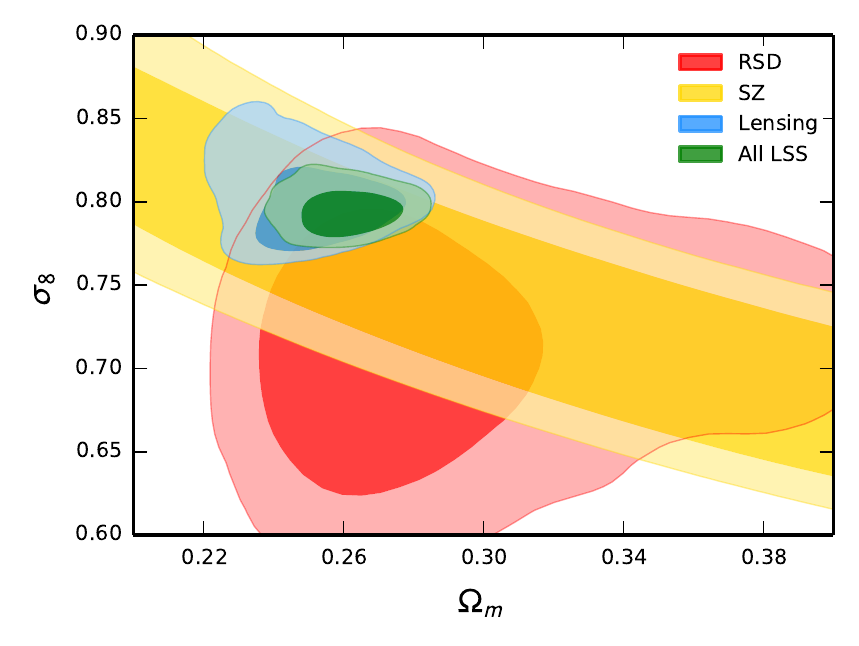}
  \caption{Constraints on the $\sigma_8-\Omega_{\rm m}$ plane for LSS data {\em only} with {\em Planck} priors on $\Theta_{\rm MC}$ and $n_{\rm S}$ to avoid over-fitting the data. 1 and $2\sigma$ contours are shown for RSD (red), SZ clusters with $1-b = [0.7,1.0]$ (gold) and lensing (blue). The joint LSS constraint (LSSall, green)  comes from combining the 3 different LSS probes.}
  \label{fig:omegam_sigma8_lss}
\end{figure}

\begin{figure}
\centering
  \subfloat{\includegraphics{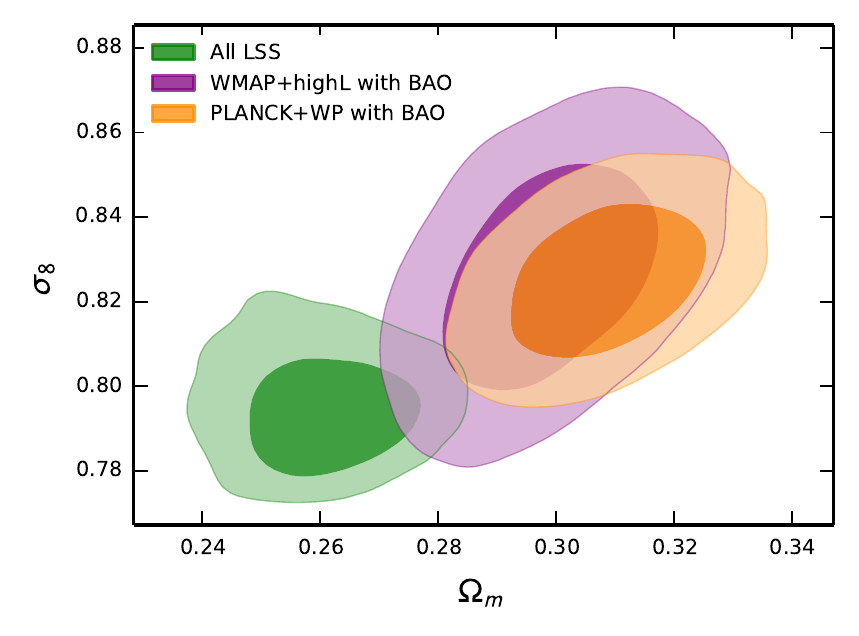}}\hskip 0cm
  \subfloat{\includegraphics{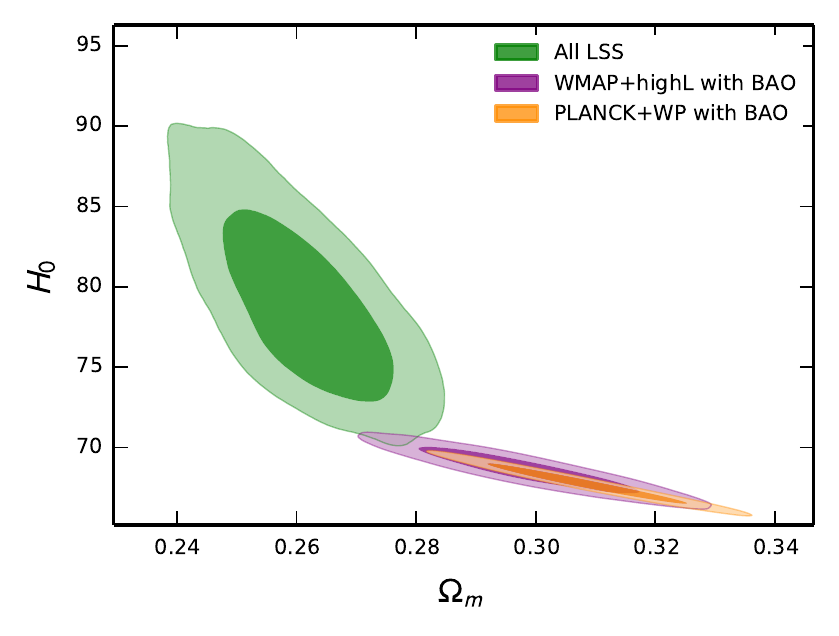}}
  \caption{The putative tension between CMB and LSS measurements. The left-hand plot shows the constraint on $\sigma_8-\Omega_{\rm m}$, where the LSSall contour (green) is as shown in Fig. \ref{fig:omegam_sigma8_lss}, and {\it Planck}+WP+BAO (orange) and WMAP+highL+BAO (purple). It is clear that there is a discrepancy with the CMB and the joint LSS constraints on this parameter combination. The right-hand plot shows the $H_0-\Omega_{\rm m}$ plane for the same data, where the tension is even more apparent.}
  \label{fig:omegam_sigma8}
\end{figure}

Next, we consider the $\sigma_8-\Omega_{\rm m}$ plane presented in Figs.~\ref{fig:omegam_sigma8_lss} and \ref{fig:omegam_sigma8}. In Fig.~\ref{fig:omegam_sigma8_lss} we show the results of constraining a 5 parameter model with parameters ${\bf p}=\{\Omega_{\rm b}h^2, \Omega_{\rm c}h^2, \Theta_{\rm MC}, A_{\rm S}, n_{\rm S}\}$ using LSS data {\em only} (note that $\tau_{\rm R}$ plays no role in LSS observables). We also note that, following~\cite{2013arXiv1303.5076P}, we fix $\sum m_{\nu} = 0.06$ eV within the $\Lambda$CDM in order to satisfy the results from oscillation experiments. Initially we use the individual LSS measurements: SZ cluster counts, with the more conservative choice $1-b=[0.7,1.0]$, lensing and RSD, in conjunction with the well determined {\it Planck} priors $\Theta_{\rm MC}=1.04131\pm0.00063$ and $n_{\rm S}=0.9603\pm0.0073$~\cite{2013arXiv1303.5076P} to avoid over-fitting the model. We then combine the three together to create a joint LSS constraint (which we will denote as ``LSSall" in subsequent sections) that yields $\sigma_8=0.7946\pm0.0094$ and $\Omega_{\rm m}=0.2610\pm0.0093$.

 Visually it is clear that the $1\sigma$ limits of the three LSS datasets are in agreement with each other, and that the joint constraints appears to be mainly driven by the lensing data. In Fig.~\ref{fig:omegam_sigma8} this is compared with constraints on the standard 6 parameter $\Lambda$CDM model from CMB data. These give $\sigma_8=0.825\pm0.012$ and $\Omega_{\rm m}=0.309\pm 0.011$ for {\it Planck}+WP+BAO and $\sigma_8=0.827\pm0.017$ and $\Omega_{\rm m}=0.299\pm0.012$ for WMAP+highL+BAO. It is clear that there is a discrepancy between the joint LSS constraint and that from CMB+BAO, that is stronger  for {\it Planck} + WP, but is still significant for WMAP+highL.
 
 Comparing the likelihoods of the difference between LSSall and \emph{Planck}+WP+BAO for the 5 parameter model allows us to quantify the degree of tension between the data, which we find to exceed $5\sigma$. To calculate this we find the mean and covariance of the 5 parameters by integrating along each dimension of the 5D parameter space. The difference in the means is found and the covariance matrices combined such that a multivariate normal distribution can be sampled to form a probability distribution function (pdf). Integrating the 5D volume inside the contour formed by taking the value of the pdf when the difference in the means is zero gives the confidence level of the amount of tension~\cite{Rencher:2003}.

There is a prima facie case that the CMB and LSS measurements discussed in this paper are in tension, at a level surpassing $5\sigma$. As presented here it is a model independent statement that the measurements of the CMB, that are dominated by large-scales, are incompatible with smaller scale measurements, quantified by $\sigma_8$, from LSS. The fact that there are three separate LSS measurements that are all mildly incompatible with the CMB, but are sufficiently compatible with each other to provide a coherent constraint on small-scales, builds confidence that this could be a real phenomenon. In subsequent sections we will investigate possible modifications to the model that would provide a resolution to this tension.

\section{Inclusion of neutrinos}
\label{sec:neutrinos}

The inclusion of a neutrino component in the cosmological model can reduce the amount of power on small scales for a given large-scale normalisation, $A_{\rm S}$. This is true both in the case of active neutrinos that correspond to the mass eigenstates of the standard three flavours and also for a sterile neutrino, which evades the strong bound on the number of neutrino species from particle physics experiments by not being involved in weak interactions. 

\subsection{Active neutrinos}
\label{sec:activeneutrinos}
The inclusion of active neutrinos is modelled by the addition of a single parameter, $\sum m_{\nu}$, assuming that this is distributed equally amongst the 3 species of massive neutrino. This approximates a degenerate hierarchy with $m_1=m_2=m_3=\sum m_{\nu}/3$, which is true for large $\sum m_{\nu}$, as is the case in the models we will find gives rise to the best fit to the data. Within the currently constrained limits, such models affect structure growth on small scales and the primary anisotropies of the CMB. A detailed description of these effects can be found in \cite{Elgaroy:2004,Lesgourgues:2006,Hannestad:2010,Hall:2012}; here, we present a brief description of the salient features.

There is little difference between massive (with $\sum m_\nu\lesssim 0.5$ eV) and massless neutrinos in terms of their effect on pre-recombination dynamics -- both the background and of perturbations -- since they are relativistic at recombination in both cases. The differences that do arise are due to the ratio of the angular diameter distance to last-scattering, $D_{\rm A}(z_*)$, to the sound horizon at last-scattering, $r_{\rm s}(z_*)$, which sets the angular scale of the CMB acoustic peaks. As the mass of the neutrino increases,  $D_{\rm A}(z_*)$ decreases,  last-scattering appears closer and anisotropies are shifted to larger angular scales \cite{Hall:2012}. There is a degeneracy in the effect on the CMB primary anisotropies between dark energy density and massive neutrinos in flat space, in addition to  a difference in the Hubble constant, but this degeneracy is broken by several effects including the late-time integrated-Sachs-Wolfe (ISW) effect (see \cite{Riess:2011,Howlett:2012,Lewis:2002} for more details). The CMB primary anisotropies are affected via the back-reaction on the metric perturbations from the stress-energy of neutrino perturbations. The size of the effect on the CMB is ${\cal O}[(\sum m_\nu/k_{\rm B} T_\nu)^2] \rho_\nu/\rho_{\text{tot}}$ where $\rho_\nu$ is the energy density per species of massless neutrino. For neutrino mass scales relevant to this analysis, changes in the CMB should be $\sim$\phantom{ }0.1\% as found in \cite{Hall:2012} where they used $\sum m_\nu=0.37$eV.

Massive neutrinos also reduce structure growth on small scales compared to massless neutrinos. Neutrinos cluster on scales above their free-streaming length -- for a non-relativistic transition in matter domination the free-streaming length is $\lambda_{\rm FS}(z)\propto a^{1/2}$ (see, for example, \cite{Lesgourgues:2006}) and therefore the comoving free-streaming length decreases with time. The growth of structure is reduced since neutrinos whose Fourier modes are inside the comoving horizon at the non-relativistic transition cannot cluster until they leave the shrinking comoving free-streaming length. There is suppression in the matter power spectrum on smaller scales due to the massive neutrinos which are currently within the comoving free-streaming length. This is found to have scale-free fractional suppression of $\sim -8f_\nu$ where $f_\nu=\Omega_\nu/\Omega_{\rm m}$~\cite{Hu:1997}.

\begin{figure}
  \includegraphics{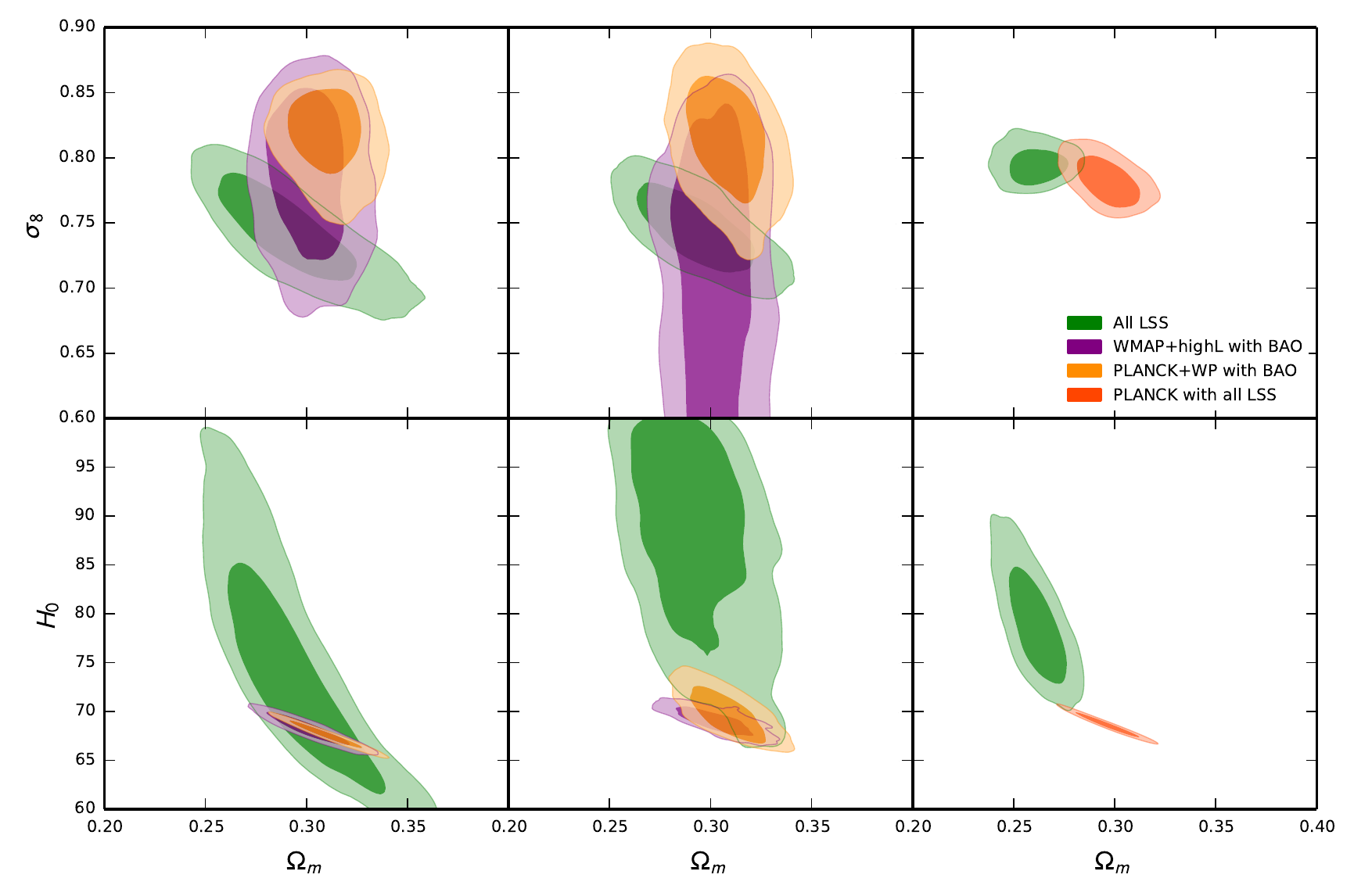}
  \caption{Each panel shows the same data as those presented in in Fig.~\ref{fig:omegam_sigma8} but with the inclusion of (i) $\sum m_{\nu}$ in the left column, and  (ii) $m_{\rm sterile}^{\rm eff}$ and $\Delta N_{\rm eff}$ in the middle column. The right hand column shows \emph{Planck} with RSD, lensing and SZ {\em without} WP (dark orange), where $\tau_{\rm R}$ is allowed to vary. The top row shows the reduction in tension in the $\sigma_8-\Omega_{\rm m}$ plane with the addition of each piece of ``new physics'', whilst the bottom row shows the $\sigma_8-H_0$ plane. It is clear that the possible values of $H_0$ allowed by LSS increases drastically when either active or sterile neutrinos are added. }
  \label{fig:omegam_sigma8_m3}
\end{figure}

We present the equivalent of the two plots in Fig.~\ref{fig:omegam_sigma8} in the left-hand panels of Fig.~\ref{fig:omegam_sigma8_m3} when $\sum m_{\nu}$ is allowed to vary. There is a significant reduction in the tension between the combined LSS constraint  (green contours) and the CMB observations from WMAP+highL+BAO (purple contours). It appears that there is still a tension in the case of  \emph{Planck}+WP+BAO (orange contours) although this is weaker than in the case when $\sum m_{\nu}$ is fixed to 0.06eV. We will quantify this residual tension below.

\begin{figure}
    \centering
    \includegraphics{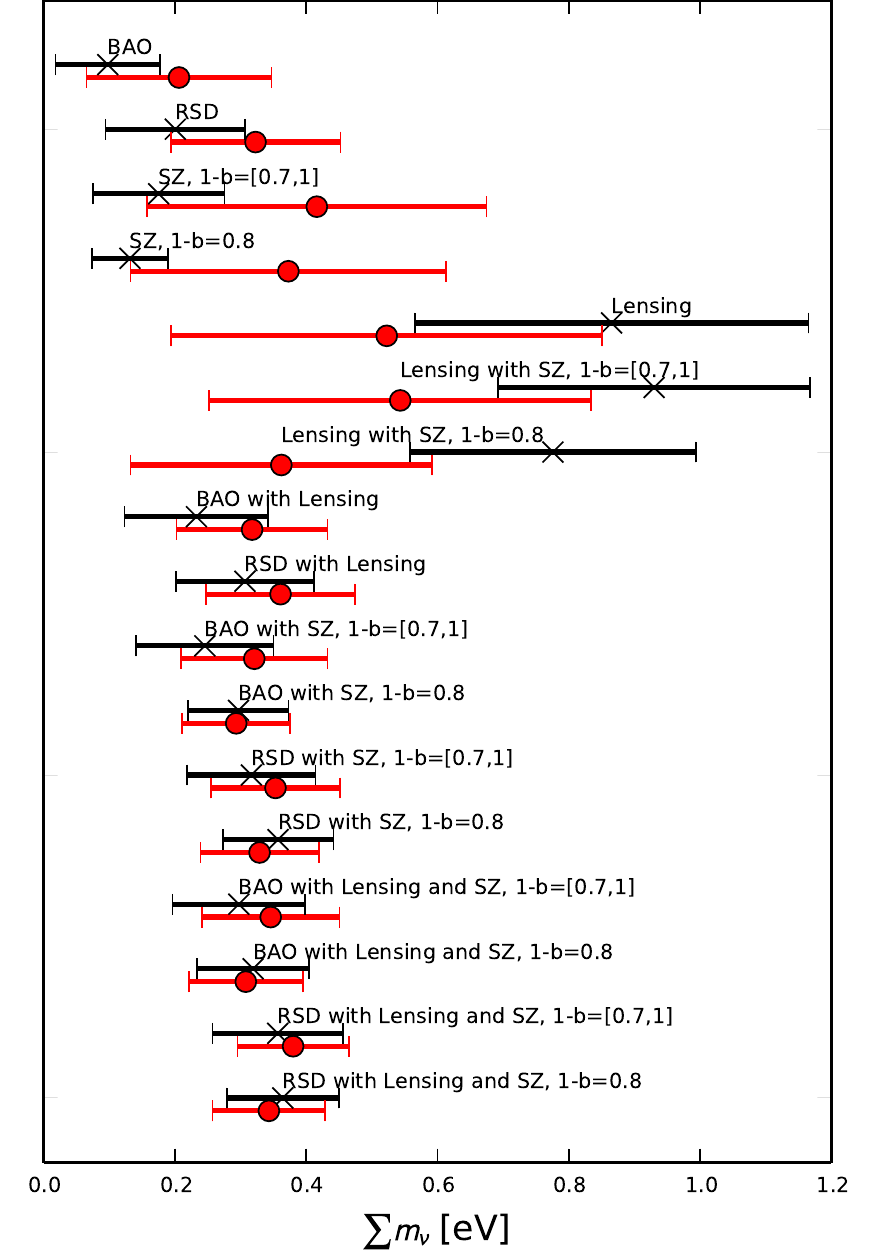}
    \caption{Comparison of the the 1D marginalised value $\sum m_{\nu}$ and $1\sigma$ errors for a wide range of LSS data combinations with CMB data. The CMB data used is \emph{Planck}+WP (black) or WMAP+highL (red). In some cases there is clearly only an upper bound, but as the number of LSS datasets included increases the constraint stabilises to a non-zero value with a significance of around $3-4\sigma$.  It is clear that preference for non-zero $\sum m_{\nu}$ is not dependent on this choice when two or more LSS datasets are included. Moreover, it is clear that there is a preference for non-zero neutrino mass {\em without} including SZ data. }
    \label{fig:comparison}
\end{figure}

Comparing Fig. \ref{fig:omegam_sigma8} to Fig. \ref{fig:omegam_sigma8_m3} suggests that inclusion of $\sum m_{\nu}$ as a parameter might resolve the discrepancy highlighted in section~\ref{sec:tension}. We have performed such analyses for a wide range of data combinations and the range of marginalised constraints on $\sum m_{\nu}$ are presented in Fig.~\ref{fig:comparison}. We studied both the case where the CMB data used comprises \emph{Planck}+WP (black) or WMAP+highL (red). As one goes down the page the number of LSS datasets included increases. When one is included there is a preference for non-zero $\sum m_{\nu}$, but the significance is low and only 95\% upper bounds can be imposed. However when two or more LSS datasets are included one can infer constraints such that  $\sum m_{\nu}>0$ at significance levels of around $3-4\sigma$. The combination of CMB+BAO+lensing+SZ (with $1-b=0.8$) was studied in \cite{Battye:2013}. The numbers presented here are compatible with those, even though we now use the 6-bin CFHTLenS data, and moreover they are compatible with CMB+BAO+lensing+RSD which {\em does  not} include any potentially uncertain measurement from SZ clusters. 

\begin{figure}
    \centering
        \includegraphics{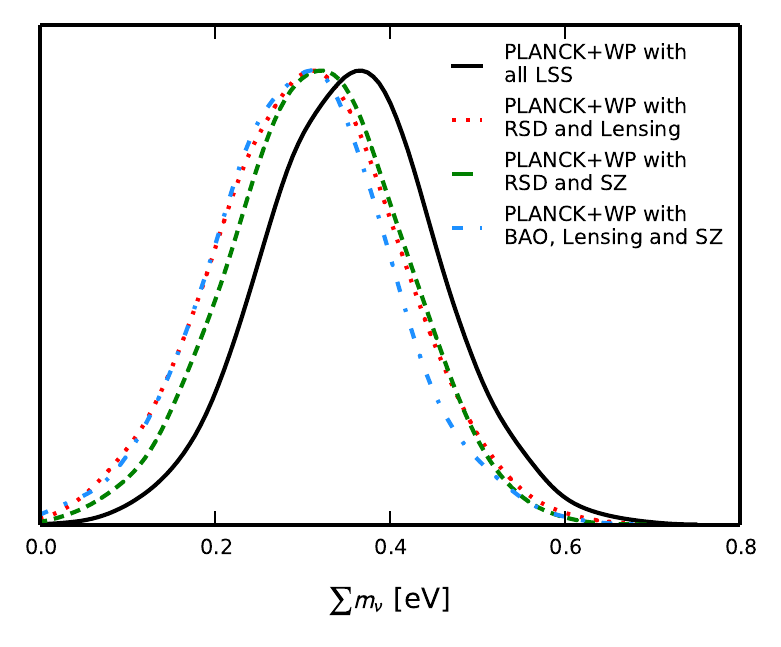}
    \caption{1D marginalised likelihoods for $\sum m_{\nu}$ for \emph{Planck}+WP and different combinations of LSS data. We show \emph{Planck}+WP+LSSall (solid black), which gives $\sum m_{\nu}=(0.357\pm 0.099)\,{\rm eV}$, and combinations of two LSS probes:  RSD+lensing (red dotted),  RSD+SZ (green dashed) and BAO+lensing+SZ (blue dot-dashed). In each case for SZ data we use the prior $1-b=[0.7,1.0]$.}
    \label{fig:likelihood}

\end{figure}

Given the three way agreement between the LSS data it seems reasonable to combine them and consider \emph{Planck}+WP+LSSall  (where LSSall uses the more conservative SZ prior of $1-b=[0.7,1.0]$) as the headline result from this paper. The 1D marginalised likelihoods for $\sum m_{\nu}$ are presented in Fig.~\ref{fig:likelihood} for \emph{Planck}+WP plus permutations of 2 LSS datasets. The full set of fitted parameters are presented in the 2nd column of Table~\ref{table:AST}. For the combination of  \emph{Planck}+WP+LSSall  we find that $\sum m_{\nu}=(0.357\pm 0.099)\,{\rm eV}$ compared to a 95\% upper limit of $\sum m_{\nu}<0.258\,{\rm eV}$ for \emph{Planck}+WP+BAO. The largest change in the other fitted parameters is a $\sim 1.6\sigma$ shift in $\Omega_{\rm c}h^2$. As expected the fitted value of $\sigma_8$ shifts from $0.818\pm 0.023$ to $0.749\pm 0.019$ when LSS is included, but the value of $\Omega_{\rm m}$ actually increases by around $1\sigma$, presumably since the massive neutrinos contribute to it.

\begin{table}
  \centering
  \begin{tabular}{|c|c|c|c|c|c|c|}\hline
    &\multicolumn{2}{c|}{Active Neutrinos}&\multicolumn{2}{c|}{Sterile Neutrinos}&\multicolumn{2}{c|}{$\tau_{\rm R}$}\\\hline
    Parameter&I&II&I&II&I&II\\\hline
    $\Omega_{\rm b}h^2$	&0.02213$\pm$0.00025	&0.02229$\pm$0.00025	&0.02237$\pm$0.00029	&0.02250$\pm$0.00028	&0.02210$\pm$0.00025	&0.02224$\pm$0.00026\\\hline
    $\Omega_{\rm c}h^2$	&0.1185$\pm$0.0019	&0.1154$\pm$0.0014	&0.1250$\pm$0.0050	&0.1180$\pm$0.0045	&0.1187$\pm$0.0018	&0.1166$\pm$0.0017\\\hline
    100$\theta_{\rm MC}$&1.04141$\pm$0.00057	&1.04156$\pm$0.00056	&1.04067$\pm$0.00073	&1.04122$\pm$0.00063	&1.04137$\pm$0.00058	&1.04143$\pm$0.00058\\\hline
    $\tau_{\rm R}$	&0.092$\pm$0.013	&0.096$\pm$0.014	&0.096$\pm$0.014	&0.096$\pm$0.015	&0.091$\pm$0.013	&0.049$\pm$0.021\\\hline
    $n_{\rm S}$		&0.9627$\pm$0.0061	&0.9677$\pm$0.0055	&0.977$\pm$0.010	&0.969$\pm$0.010	&0.9618$\pm$0.0058	&0.9635$\pm$0.0061\\\hline
    $\log(10^{10}A_{\rm S})$&3.090$\pm$0.025	&3.091$\pm$0.027	&3.115$\pm$0.029	&3.105$\pm$0.030	&3.088$\pm$0.025	&3.000$\pm$0.039\\\hline\hline
    $\sum m_{\nu}$[eV]	&$<$0.26		&0.357$\pm$0.099	&			& 			& 			& \\\hline
    $m_{\nu{\rm, sterile}}^{\rm eff}$[eV]& 	& 			&$<$0.48		&0.66$\pm$0.18		& 			& \\\hline
    $\Delta N_{\rm eff}$& 			& 			&0.47$\pm$0.27		&0.32$\pm$0.21		&			& \\\hline\hline
    $H_{0}$		&67.57$\pm$0.92		&66.52$\pm$1.15		&69.79$\pm$1.71		&68.00$\pm$1.11		&67.73$\pm$0.80		&68.64$\pm$0.80\\\hline
    $\Omega_{\rm m}$	&0.311$\pm$0.012	&0.320$\pm$0.015	&0.308$\pm$0.012	&0.321$\pm$0.013	&0.309$\pm$0.011	&0.296$\pm$0.010\\\hline
    $\sigma_8$		&0.818$\pm$0.023	&0.749$\pm$0.019	&0.817$\pm$0.030	&0.736$\pm$0.017	&0.825$\pm$0.012	&0.783$\pm$0.012\\\hline
    $z_{\rm re}$	&11.21$\pm$1.11		&11.15$\pm$1.20		&11.77$\pm$1.18		&11.66$\pm$1.25		&11.13$\pm$1.08		&6.91$\pm$2.20\\\hline\hline
    $-2\ln\mathcal{L}_{\it Planck}$&7791.00	&7796.95		&7790.19		&7793.165		&7791.44		&7797.55\\\hline
    $-2\ln\mathcal{L}_{\rm WP}$&2014.26		&2014.93		&2015.18		&2014.47 		&2014.26		&{\color{red}2029.00*}\\\hline
    $-2\ln\mathcal{L}_{\rm BAO}$&1.39		&{\color{red}5.69*}	&1.49			&{\color{red}6.94*}	&1.40			&{\color{red}1.53*}\\\hline
    $-2\ln\mathcal{L}_{\rm RSD}$&{\color{red}-12.55*} &-14.80		&{\color{red}-10.28*} 	&-13.66			&{\color{red}-11.90*} 	&-15.05\\\hline
    $-2\ln\mathcal{L}_{\rm Lensing}$&{\color{red}-4506.57*}&-4523.11	&{\color{red}-4494.89*} &-4530.10		&{\color{red}-4500.15*}	&-4519.93\\\hline
    $-2\ln\mathcal{L}_{\rm SZ}$	&{\color{red}12.86*}&0.34 		&{\color{red}19.686*}	&0.19			&{\color{red}15.21*}	&2.05\\\hline
    $-2\ln\mathcal{L}$		&9806.65	&5274.30		&9806.86		&5264.06		&9807.09		&3264.71\\\hline
  \end{tabular}
\caption{Marginalised parameter table for scenarios discussed in the text. Data combinations are I = \emph{Planck}+WP+BAO and  II = \emph{Planck}+WP+LSSall. For the $\tau_{\rm R}$ runs then WP is omitted in column II. The numbers in red and denoted * are the values of likelihoods that are not included in the fit calculated, but show the level of tension. Note that when we use data combination II the values of $\sigma_8$ are smaller than those using I, but those for $\Omega_{\rm m}$ are increased in both the active and sterile neutrino cases.}
\label{table:AST}
\end{table}

The high values of $\sum m_{\nu}$ are not favoured by the \emph{Planck}+WP+BAO data: values of $\sum m_{\nu}$ for \emph{Planck}+WP+LSSall are in tension with the upper limit from \emph{Planck}+WP+BAO. As discussed in \cite{Battye:2013}, this can be quantified by performing the analysis using two different neutrino masses and considering the statistics of the difference. To be concrete, in the MCMC analysis we assume two masses, $\sum m_{\nu}^{\rm CMB}$ and $\sum m_{\nu}^{\rm LSS}$, and calculate all observables (CMB power spectra, lensing convergence etc.) for each.  For any CMB data we use the observables calculated using $\sum m_{\nu}^{\rm CMB}$, and for any LSS data we use the observables using $\sum m_{\nu}^{\rm LSS}$. The other cosmological parameters are shared. We can then compute the marginalised posterior $\Delta M=\sum m_{\nu}^{\rm CMB}-\sum m_{\nu}^{\rm LSS}$ and quantify at what significance this is non-zero.  We have performed such an analysis and find  $\Delta M = 0.32 \pm 0.13$ eV, i.e. non-zero at the $2.5 \sigma$ level. This quantifies the extent to which the active neutrino model is in tension with {\em Planck data}.

\subsection{Sterile neutrinos}

There are a host of anomalies from short baseline neutrino oscillation experiments which may be solved by the addition of a sterile neutrino. Firstly the LSND experiment~\cite{LSND} observes an excess of $\bar{\nu}_e$ candidates, suggesting the oscillation $\bar{\nu}_\mu\to\nu_s\to\bar{\nu}_e$ where the mass of the sterile neutrino is constrained to $\sim 1$ eV  by the KARMEN~\cite{KARMEN} and Bugey~\cite{Bugey} experiments. The MiniBooNE experiment~\cite{MiniBooNE}, as well as testing the LSND signal, also detects an excess of $\nu_e$ from the neutrino mode rather than anti-neutrino mode. Although the neutrino mode does not completely agree with the expected sterile neutrino signal, there are several explanations due to the method of detection~\cite{Katori:2014}. Reactor anomalies detect a 6\% lower rate of electron anti-neutrinos than is expected, which can be interpreted as neutrino oscillations with a 1 eV sterile neutrino~\cite{Mention:2011}. Lower event rates of $\nu_e+ ^{71}{\rm Ga}\to ^{71}{\rm Ge}+e^-$ than expected can also be explained by 1 eV sterile neutrino oscillations, solving what is known as the Gallium anomaly~\cite{Giunti:2010}.

Recently, joint analysis using cosmological and short baseline data has been carried out for models with both one and two added sterile neutrinos~\cite{Archidiacono:2013}. The addition of short baseline data in the form of priors on the cosmological data lowers the mass of the sterile neutrino, in a single sterile neutrino model, to the $m_{\rm sterile}\sim1$ eV range at high significance. A model with two added sterile neutrinos is generally not allowed since this leads to a universe with too much radiation~\cite[pg. 163]{Les:nc}.

The standard approach to modelling sterile neutrinos is to introduce two new parameters into the fitting process. The first is an effective neutrino mass, $m_{\rm sterile}^{\rm eff}$, and the second is the change in  the effective number of degrees of freedom, $\Delta N_{\rm eff}$, such that $N_{\rm eff}=3.046+\Delta N_{\rm eff}$. In this case the active neutrinos are modelled as 1 massive neutrino with $m_{\nu}=0.06\,{\rm eV}$ and 2 massless neutrinos, which would accurately model a normal hierarchy with $m_1 < m_2\ll m_3$. The cosmological results are not very sensitive to this assumption on the structure of the neutrino active sector. The density of sterile neutrinos is given by 
\begin{equation}
\Omega_{\rm sterile}h^2={m^{\rm eff}_{\rm sterile}\over 94\,{\rm eV}}\,.
\end{equation}
In the MCMC analysis we impose a prior $m_{\rm sterile}^{\rm eff}/\Delta N_{\rm eff}< 10\,{\rm eV}$, as used in the \emph{Planck} analysis~\cite{2013arXiv1303.5076P}, since sterile neutrinos with large effective masses become degenerate with cold dark matter. 

This parameterisation encompasses a wide of range of possible models for sterile neutrinos, which are typically formed in the early Universe by oscillations. Two possible scenarios that have been widely discussed in the literature are: 

\begin{enumerate}
\item ``On resonance'' oscillations in the Dodelson-Widrow model~\cite{Dodelson:1993}. In this case the sterile neutrinos have the same temperature as their active counterparts and are formed via oscillations when there is no lepton asymmetry and the mixing angle is small. Under the assumption that neutrino decoupling is instantaneous, the distribution function for the neutrinos is 
\begin{equation}
f_{\rm DW}(p)=\frac{\Delta N_{\rm eff}}{\exp[p/T_\nu]+1}\,.
\end{equation}
In this case $\Delta N_{\rm eff}$ is a constant, which is not necessarily an integer, and the true mass is given by $m_{\rm sterile}=m_{\rm sterile}^{\rm eff}/\Delta N_{\rm eff}$.

\item ``Off resonance'' oscillations leading to a thermal scenario~\cite{Shi:1998}. The distribution function is standard Fermi-Dirac with thermal temperature $T_{\rm sterile}$,  given by  
\begin{equation}
  T_{\rm sterile}=\Delta N_{\rm eff}^{1/4} T_{\nu}\,,
  \end{equation}
  where $T_{\nu}$ is the thermal temperature of the active neutrinos. $\Delta N_{\rm eff}$, again not necessarily an integer, quantifies the level of thermalisation with $\Delta N_{\rm eff}=1$ corresponding to the complete thermalisation, for one species of sterile neutrino, with $T_{\rm sterile}=T_{\nu}$. The thermal mass of the neutrinos is given by 
\begin{equation}
m_{\rm sterile}=\Delta N_{\rm eff}^{-3/4}m_{\rm sterile}^{\rm eff}\,.
\end{equation}

\end{enumerate}

In the previous section on active neutrinos we performed a detailed analysis of a wide range of LSS data combined with CMB data from \emph{Planck}+WP or WMAP+highL. We found that the combination of CMB and 2 or more LSS datasets leads to consistent conclusions. We find similar results for the sterile neutrino case, so when presenting the results we will restrict our attention to this limited range of possibilities. 

\begin{figure}
  \centering
  \subfloat{\includegraphics{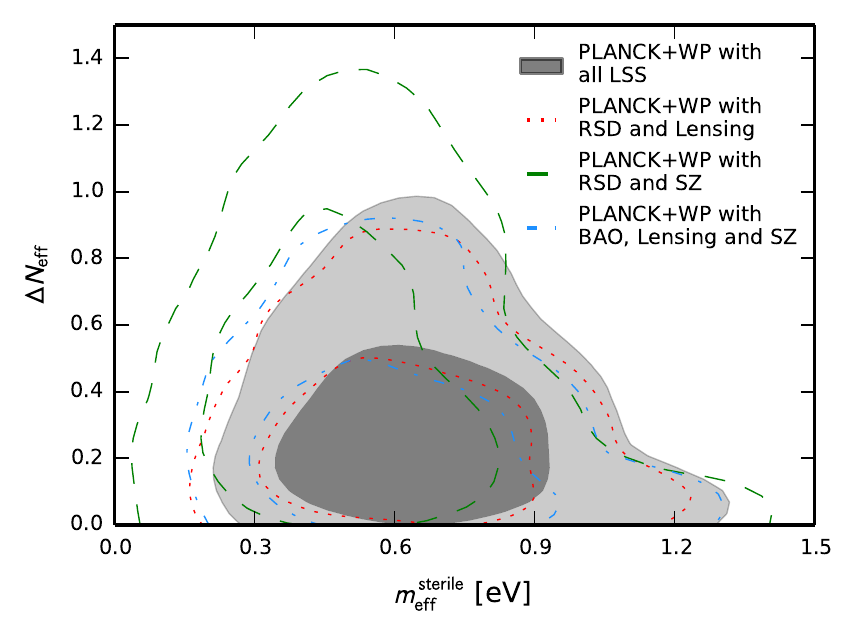}}\hskip0cm
  \subfloat{\includegraphics{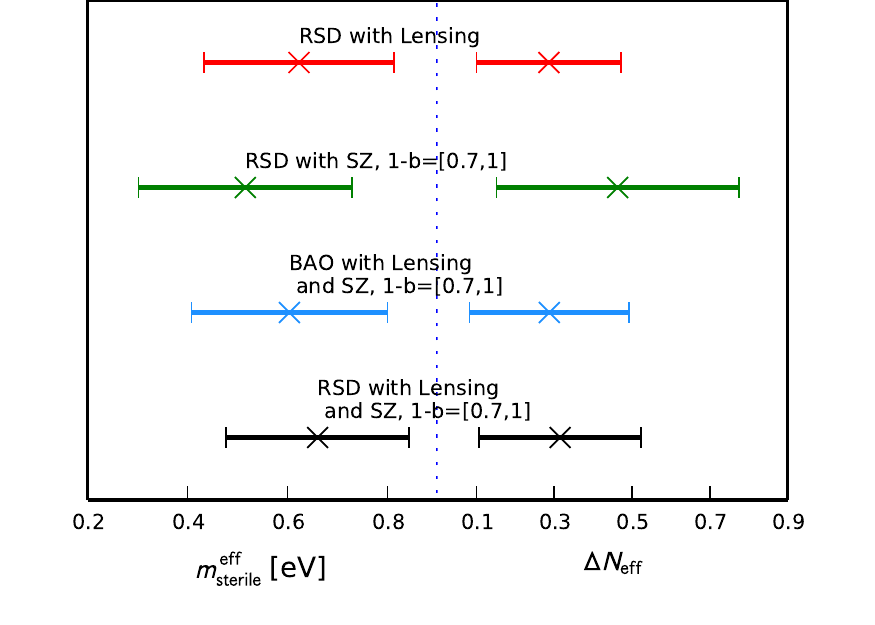}}
  \caption{(Left) 2D Likelihood plots for $m_{\rm eff}^{\rm sterile}$ and $\Delta N_{\rm eff}$ for different data combinations:  \emph{Planck}+WP+LSSall (black), \emph{Planck}+WP+RSD+lensing (red dotted),  \emph{Planck}+WP+RSD+SZ (1-b=[0.7,1]) (green dashed) and \emph{Planck}+WP+BAO+lensing +SZ clusters (1-b=[0.7,1]) (blue dot-dashed). (Right) Marginalised 1D values of  $m^{\rm eff}_{\rm sterile}$ and $\Delta N_{\rm eff}$  for different data sets using the same colour scheme, showing the mean value and $1\sigma$ errors.}
  \label{fig:meffster}
\end{figure}

The middle panel of Fig.~\ref{fig:omegam_sigma8_m3} illustrates the impact of including $m_{\rm sterile}^{\rm eff}$ and $\Delta N_{\rm eff}$ as parameters on the constraints in the $\sigma_8-\Omega_{\rm m}$ and $H_0-\Omega_{\rm m}$ plane for various data combinations. As with the case of active neutrinos there is clear evidence that inclusion of sterile neutrinos can ameliorate the discrepancy highlighted in section~\ref{sec:tension}. In Fig.~\ref{fig:meffster} we present the results of joint CMB and LSS analyses. There is a consistent picture for a range of combinations of these data suggesting a non-zero value for $m_{\rm sterile}^{\rm eff}$. Using \emph{Planck}+WP+LSSall we find that $m_{\rm sterile}^{\rm eff}=(0.67\pm 0.18)\,{\rm eV}$ and $\Delta N_{\rm eff}=0.32\pm 0.20$ -- marginalised parameters are presented in Table~\ref{table:AST}. This corresponds to a significant preference for the sterile neutrino model using the joint likelihood.

It was pointed out in \cite{Wyman:2013,Battye:2013} that the extra degree of freedom due to $\Delta N_{\rm eff}$ allowed for a best-fitting value of $H_0$ more compatible with low redshift measurements, for example, using Cepheids~\cite{Riess:2011yx}. This is due to the degeneracy between $\Delta N_{\rm eff}$ and $H_0$. However, with the inclusions of RSDs an even lower value of $\sigma_8$ is preferred (see Fig.~\ref{fig:omegam_sigma8_lss}). Due to the $\sigma_8-H_0$ degeneracy the result is a lower value of $H_0$ more closely aligned with the {\it Planck}+WP value. 

As with active neutrinos, the improvement in the likelihood when including the extra parameters comes at a price;  the \emph{Planck} component of the likelihood is increased by $\Delta \chi^2\approx 4$. This is less problematic than in the case of  active neutrinos since the extra freedom from $N_{\rm eff}$ allows a better fit to the CMB, but it is still unsatisfactory.  We perform a similar analysis to quantify the residual level of tension. In the MCMC we define CMB parameters,  $m_{\rm sterile}^{\rm eff, CMB}$ and   $\Delta N_{\rm eff}^{\rm CMB}$, and LSS parameters  $m_{\rm sterile}^{\rm eff, LSS}$ and   $\Delta N_{\rm eff}^{\rm LSS}$, with all other parameters shared. As before, each CMB or LSS observable uses the relevant neutrino parameters. The joint marginalised distribution of $m_{\rm sterile}^{\rm eff,  LSS}- m_{\rm sterile}^{\rm eff, CMB}$ and  $\Delta N_{\rm eff}^{\rm LSS}-\Delta N_{\rm eff}^{\rm CMB}$ is then constructed, and it is found that both parameters are non-zero at the $1.6\sigma$ level. 

\section{Alternative explanations}
\label{sec:alternative}

In the previous sections we have first made the statement that there is a strong case for a tension between CMB and LSS measurements, and we have studied in detail their possible resolution using massive neutrinos, both active and sterile. An obvious question that arises, particularly since we have already pointed out that massive neutrinos only partially resolve the discrepancy: is there a better modification to the standard cosmological model that might accommodate the two types of data? In this section we make a non-exhaustive survey of other possible explanations.

\subsection{Ignore WMAP polarization data}

The measurement of the E-mode polarisation on large-scales by WMAP is crucial in all the previous analyses. The CMB temperature anisotropies constrain the parameter combination $A_{\rm S}e^{-2\tau_{\rm R}}$ in the absence of the ISW effect and this requires a measurement of polarisation on large scales to infer  $\tau_{\rm R}$ and hence allow $A_{\rm S}$ to be deduced independently.  The small-scale amplitude $\sigma_8$ is a derived parameter and is sensitive to all the cosmological parameters, but it is $\propto \sqrt{A_{\rm S}}$. If $\tau_{\rm R}$ were lower than the $\tau_{\rm R}=0.091 \pm 0.013 $ as required by \emph{Planck}+WP then $\sigma_8$ would reduce $\propto e^{\tau_{\rm R}}$. In particular a reduction of $\sigma_8$ from $\approx 0.83$ as suggested by CMB measurements to $\approx 0.78$, which is closer to the value preferred by the LSS measurements,  would require $\tau_{\rm R}$ to be reduce from $\approx 0.09$ to $\approx 0.05$. Of course, this would require the WMAP polarisation measurement on large scales to have been misinterpreted. However, this is the regime where instrumental systematics and foreground subtraction are most difficult and therefore it seems at least sensible to consider such a possibility. 

\begin{figure}
  \centering
  \includegraphics{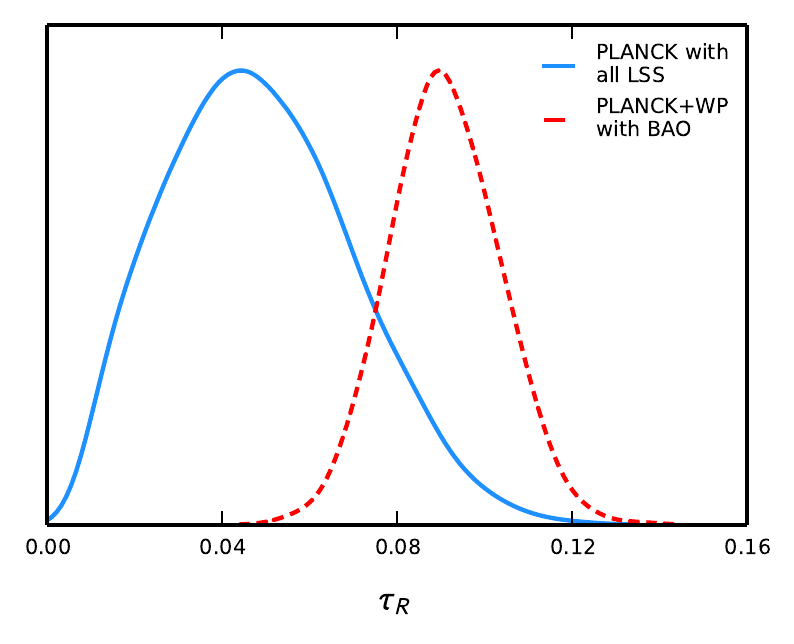}
  \caption{Marginalised likelihood distributions for $\tau_{\rm R}$ within the standard $\Lambda$CDM model, for \emph{Planck}+LSSall (solid blue) and \emph{Planck}+WP+BAO (dashed red). Using WP to constrain $\tau_{\rm R}$ leads to a narrow distribution of values centred on 0.09 whereas using the LSS data favours a much lower value, albeit with a wider distribution.}
  
  \label{fig:tau}
\end{figure}

For the moment, in order to illustrate the point that the LSS measurements can be used to fix $\tau_{\rm R}$ in the absence of the a large scale polarisation measurement, we have removed WP from the likelihood and fitted the standard $\Lambda$CDM model to the \emph{Planck}+LSSall data. The results of doing this are presented in the final column of  Table~\ref{table:AST} and can be compared to the same using \emph{Planck}+WP data in the adjacent, 5th column. The marginalised distributions for $\tau_{\rm R}$ are presented in Fig.~\ref{fig:tau}. For \emph{Planck}+WP we find narrow range of  values of $\tau_{\rm R}$, but for \emph{Planck}+LSSall the likelihood distribution for $\tau_{\rm R}$ is much wider and $\tau_{\rm R}=0.049\pm 0.021$. As with the both the active and sterile neutrino case, the improved fit to the LSS data leads to a degradation in the fit to the \emph{Planck} temperature data quantified by $\Delta\chi^2\approx 6$. It is clear that the \emph{Planck} data has a preference for $\tau_{\rm R}\sim 0.1$ but this is not sufficiently strong to prevent the LSS dragging it to lower values in order reduce $\sigma_8$.

The lower value of $\tau_{\rm R}$ corresponds to a redshift reionization of $z_{\rm re}=6.91\pm 2.20$ for \emph{Planck}+LSSall, much lower than the value of $\approx 11$ preferred by \emph{Planck}+WP. However, such values might be considered desirable in the context of astrophysical constraints on reionization. Light from quasars show Lyman-$\alpha$ absorption, due to neutral hydrogen in the intergalactic medium, at a range of frequencies depending on the redshift. The Lyman-$\alpha$ forest is more greatly populated for larger redshift quasars, but at $z\sim 6$ all electromagnetic radiation flux below the Lyman-$\alpha$ forest drops to zero, known as the Gunn-Peterson trough~\cite{Gunn1965}. This effect is due to a large fraction of neutral hydrogen, hence indicating the boundary at the end of reionization. A complete Gunn-Peterson trough can be seen in objects such as a quasar at $z=6.28$~\cite{Becker}. Objects at even lower redshifts, $z\sim5.5$, are seen to have partial Gunn-Peterson troughs suggesting that the end of reionization was patchy~\cite{Djorgovski:2001}. These redshifts are much lower than the predicted redshifts for the beginning of reionization from WMAP polarisation data. Reionization is generally modelled as a step in the ionisation fraction, which must be a double step when taking the beginning and end of reionization at different redshifts~\cite{Kaplinghat,Kogut}, but the lower value of $\tau_{\rm R} \sim 0.05$ allows for a single step or a smooth transition from beginning to end. 

We again quantify the residual tension between the parameter value preferred by LSS compared to {\em Planck} by introducing $\tau_{\rm R}^{\rm CMB, TT}$ and $\tau_{\rm R}^{\rm CMB, TE}$ in an MCMC fit. Here  $\tau_{\rm R}^{\rm CMB, TT}$  is used for {\em Planck} temperature data and  $\tau_{\rm R}^{\rm CMB, TE}$  for large-angle WMAP polarization sourced at $z\lesssim20$. We find the marginalised posterior of the difference,  $\tau_{\rm R}^{\rm CMB, TE}-\tau_{\rm R}^{\rm CMB, TT}$, is non-zero at $2.1\sigma$. Therefore, although {\em Planck} temperature data is more compatible with the $\tau_{\rm R}^{\rm CMB, TE}$ than the active neutrino equivalent, one should also bear in mind the combined fit to CMB+LSS data of the varying $\tau_{\rm R}$ model is slightly worse.  

\subsection{Modifications to the primordial power spectrum}

The inclusion of massive neutrinos reduces the amount of small-scale power relative to large-scales in the observed matter power spectrum. One obvious possibility that needs to be considered is whether such an effect can be created using a simple modification to the initial power spectrum of adiabatic perturbations, $P_{\rm i}(k)$. Within the $\Lambda$CDM model this is $P_{\rm i}(k)=P_{\Lambda\rm CDM}=A_{\rm S}(k/k_{\rm pivot})^{n_{\rm S}}$ where we fix $k_{\rm pivot}=0.05 \, {\rm Mpc}^{-1}$. Similar endeavours have been made previously in the literature, as in \cite{Elgaroy:2001}, where the primordial power spectrum is modified via the application of a set of ``top-hat'' steps or a ``sawtooth'' shaped function to the original power spectrum. This was done as an attempt to explain specific features which could not easily be explained by a power law, $P_i(k)\propto k^n$, such as a bump-like feature in the CMB at $k\sim0.004\,h\,{\rm Mpc}^{-1}$~\cite{Griffiths:2001,Hannestad:2001}, a step-like feature between $k\sim0.06-0.6\,h\,{\rm Mpc}^{-1}$~\cite{Barriga:2001} and a dip at $k\sim0.1\,h\,{\rm Mpc}^{-1}$~\cite{Gramann:2001}. Inflationary features can be included by modifying the primordial power spectrum. Step-like features in particular can be caused by interacting scalar fields which, in turn, cause localised oscillations in the CMB which can provide a better fit to data than a featureless power spectrum~\cite{Kaiser:2012,Benetti:2013a,Benetti:2013b}. 

We will consider modifications to the primordial power spectrum that can mimic the effects of including massive neutrinos. In particular, the specific form that we will use is 
\begin{equation}
P_{\rm i}(k)=\left[1-\frac{\alpha}{2}\left(1+\tanh\left(\frac{\log \beta k}{\log \delta}\right)\right)\right]P_{\Lambda\rm CDM}(k)\,,
\end{equation}
where  $\alpha$ determines the magnitude of the overall suppression and $\beta$ and $\delta$ control the position and rate of the turn over, respectively.  For example, we find that  $\alpha=0.14$, $\beta=20\,{\rm Mpc}$   and $\delta=5$ mimics the matter power spectrum of an active neutrino model with $\sum m_{\nu}=0.3\,{\rm eV}$. Note that having the same observed matter power spectrum does not imply that the CMB power spectrum will be the same. A similar form of modification to the primordial power spectrum was proposed in \cite{Hazra:2013}, which examined how binning the primordial spectrum can produce features similar to Starobinsky inflation~\cite{Starobinsky:1992}. They find that a sharp transition is equally as probable as a smooth transition, since $\delta$ is unconstrained from below.

\begin{figure}
  \subfloat{\includegraphics{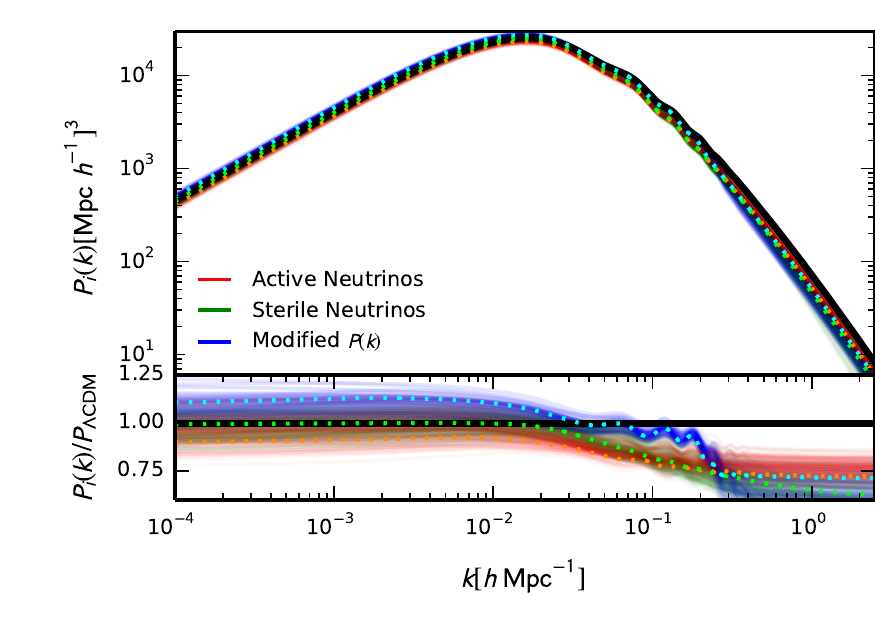}}\hskip 0cm
  \subfloat{\includegraphics{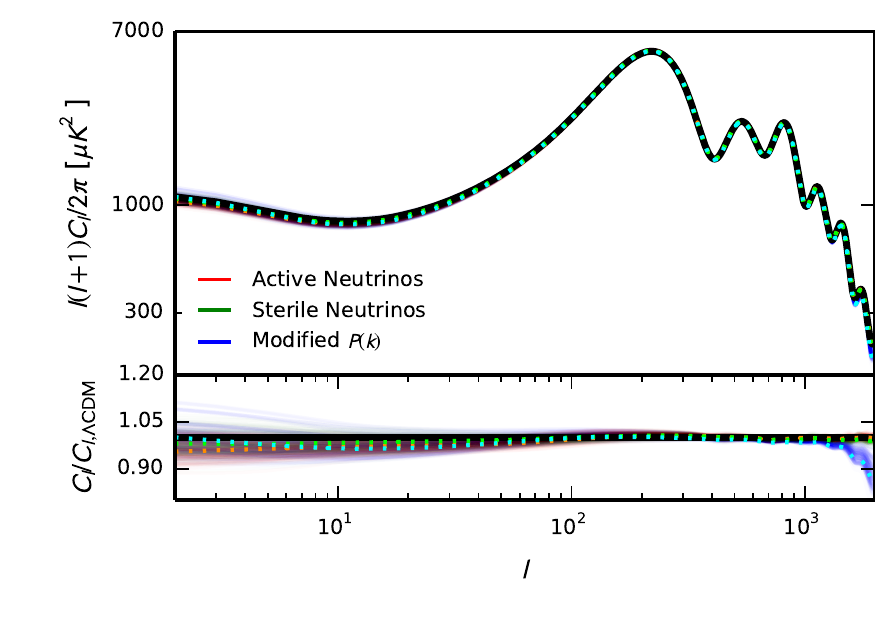}}
  \caption{(Left) Predicted matter power spectra for models colour coded by their fits to likelihoods: (i) active neutrinos (red);  (ii) sterile neutrinos (green); (iii) modified $P(k)$ (blue). (Right) Equivalent for the CMB temperature power spectrum. The overlayed dotted lines show the best fit results for each model. Power is reduced on small scales compared to large scales for each of the modifications to $\Lambda$CDM, but the modified primordial power spectrum boosts power on large scales compared to $\Lambda$CDM, whilst active neutrinos reduces it. The $C_l$s for the modified primordial power spectrum and active neutrinos differ from $\Lambda$CDM at low multipoles by around 5\%, and the modified $P(k)$ model also dips at around $\ell=1800$, causing a mildly better fit to the \emph{Planck} data than $\Lambda$CDM.}
\label{fig:Pk}
\end{figure}

After marginalisation we find $\alpha=0.32\pm0.11$, $\beta=5.96\pm0.70~{\rm Mpc}$ and $\delta=1.24\pm0.11$ with best-fits $\alpha=0.20$, $\beta=6.76~{\rm Mpc}$ and $\delta=1.12$. The combined likelihood is improved in comparison to the addition of sterile and active neutrinos by $\Delta\chi^2\approx1$ and by $\Delta\chi^2\approx11$ respectively, which can be attributed to a better fit to {\it Planck} temperature data. The {\it Planck} data has a ``dip'' at around $\ell=1800$ which corresponds with a residual systematic feature due to incomplete 4K line removal~\cite{2013arXiv1303.5076P}. It is possible the modified $P(k)$ model fits this feature better, as seen by the reduction in power around  $\ell=1800$ in the right-hand panel of Fig.~\ref{fig:Pk}.  The modified $P(k)$ model fits LSS data as well as the neutrino models. 

Given the fact there is a reduction in power for $\ell \gtrsim 2000$, we have also performed a run with {\rm Planck}+WP+LSSall+highL. We find the shape of the $\tanh$ function remains similar with ACT+SPT data, $\beta=6.00\pm2.51{\rm Mpc}$ and $\delta=0.92\pm0.48$, but the amplitude, $\alpha=0.111\pm0.083$, is tightly constrained. Therefore, it appears that this model can be excluded on the basis of a poor fit to very small scale CMB data.

\section{Discussion} \label{sec:discussion}

\subsection{Varying the lensing parameter, $A_{\rm L}$}

Weak lensing has two effects on the CMB: the first is an additional contribution to the angular power spectrum, the second is a non-Gaussian effect that has an impact on the higher-order moments. This latter effect is used in the reconstruction that had already been part of our analysis. One odd effect that has been documented in the \emph{Planck} analysis (the result being more significant when including highL data) is that, when one adds a phenomenological parameter such that $C_\ell^\Psi\to A_{\rm L}C_\ell^\Psi$ \cite{Calabrese:2008}, with $A_{\rm L}=0$ corresponding to an unlensed model and $A_{\rm L}=1$ the physical result, one finds  $A_{\rm L}=1.23\pm0.11$ \cite{2013arXiv1303.5076P} (\emph{Planck}+WP+highL) which is $\sim 2\sigma$ away from the expected value.  The reasoning behind this high value of $A_{\rm L}$ being favoured is a mystery.

We have investigated the impact of  varying of $A_{\rm L}$ on the models considered in Table~\ref{table:AST} and the equivalent results are presented in Table~\ref{table:ASTA}. We see that larger values of $\sum m_{\nu}$ and $m_{\rm sterile}^{\rm eff}$ are allowed for {\it Planck}+WP due to the degeneracy with  $A_{\rm L}$. However,  when including LSSall, the fit to the {\it Planck} component of the likelihood is still degraded, although this is less severe for active neutrinos than  when $A_{\rm L}=1$. With LSSall, the significance of the active neutrino result increases to $\sum m_{\nu}= (0.420\pm0.097)\,{\rm eV}$, but for a sterile neutrino the mass decreases, such that $m_{\rm sterile}^{\rm eff}= (0.56\pm0.15)\,{\rm eV}$. The reason for this is the correlation between the effective sterile neutrino mass and several of the observable parameters. For example, $H_0$ anti-correlates with $m_{\rm sterile}^{\rm eff}$ and since $H_0$ is larger when $A_{\rm L}$ is allowed to vary, this would suggest a lower sterile neutrino mass. $\Omega_{\rm m}$ and $z_{\rm re}$ are both correlated with $m_{\rm sterile}^{\rm eff}$ and since both $\Omega_{\rm m}$ and $z_{\rm rei}$ have lower values when $A_{\rm L}$ is varied, then this also corresponds with $m_{\rm sterile}^{\rm eff}$ being smaller.

\begin{table}
  \centering
  \begin{tabular}{|c|c|c|c|c|c|c|}\hline
    &\multicolumn{2}{c|}{Active Neutrinos}&\multicolumn{2}{c|}{Sterile Neutrinos}&\multicolumn{2}{c|}{$\tau$}\\\hline
    Parameter&I&II&I&II&I&II\\\hline
    $\Omega_{\rm b}h^2$		&0.02247$\pm$0.00030&0.02244$\pm$0.00024&0.02286$\pm$0.00037&0.02277$\pm$0.00030&0.02235$\pm$0.00028&0.02236$\pm$0.00025\\\hline
    $\Omega_{\rm c}h^2$		&0.1160$\pm$0.0023&0.1141$\pm$0.00142&0.1232$\pm$0.0052&0.1213$\pm$0.0043&0.1177$\pm$0.0019&0.1160$\pm$0.0014\\\hline
    100$\theta_{\rm MC}$	&1.04174$\pm$0.00063&1.04179$\pm$0.00055&1.0410$\pm$0.00075&1.0410$\pm$0.00066&1.04159$\pm$0.00059&1.04160$\pm$0.00057	\\\hline
    $\tau_{\rm R}$		&0.087$\pm$0.013&0.089$\pm$0.012&0.092$\pm$0.014&0.090$\pm$0.013&0.086$\pm$0.013&0.023$\pm$0.012\\\hline
    $n_{\rm S}$			&0.9696$\pm$0.0070&0.9718$\pm$0.0056&0.985$\pm$0.012&0.982$\pm$0.012 &0.9657$\pm$0.0061&0.9660$\pm$0.0054\\\hline
    $\log(10^{10}A_{\rm S})$	&3.075$\pm$0.025&3.073$\pm$0.023&3.103$\pm$0.030&3.095$\pm$0.026&3.077$\pm$0.025&2.955$\pm$0.023\\\hline
    $A_{\rm L}$		&1.28$\pm$0.13&1.182$\pm$0.055&1.29$\pm$0.13&1.137$\pm$0.063&1.19$\pm$0.10&1.202$\pm$0.055\\\hline\hline
    $\sum m_{\nu}$[eV]		&$<$0.47&0.420$\pm$0.097& & & & \\\hline
    $m_{\nu{\rm, sterile}}^{\rm eff}$[eV]& & &$<$0.82&0.56$\pm$0.15& & \\\hline
    $\Delta N_{\rm eff}$	& & &0.55$\pm$0.27&0.54$\pm$0.26& &	\\\hline\hline
    $H_{0}$		&67.75$\pm$1.04&66.58$\pm$1.14&70.59$\pm$1.71&69.66$\pm$0.15&68.37$\pm$0.89&69.02$\pm$0.67\\\hline
    $\Omega_{\rm m}$	&0.307$\pm$0.013&0.319$\pm$0.014&0.302$\pm$0.012&0.311$\pm$0.013&0.301$\pm$0.011&0.292$\pm$0.008\\\hline
    $\sigma_8$		&0.777$\pm$0.038&0.725$\pm$0.019&0.771$\pm$0.044&0.732$\pm$0.017&0.817$\pm$0.012&0.7605$\pm$0.0086\\\hline
    $z_{\rm re}$	&10.65$\pm$1.08&10.86$\pm$1.05&11.23$\pm$1.21&11.12$\pm$1.10&10.60$\pm$1.11&3.99$\pm$1.47\\\hline\hline
    $-2\ln\mathcal{L}_{\it Planck}$&7788.43&7792.43&7786.75&7792.07&7786.40&7789.41\\\hline
    $-2\ln\mathcal{L}_{\rm WP}$&2014.23&2014.69&2014.49&2014.50&2014.41&{\color{red}2047.60*} \\\hline
    $-2\ln\mathcal{L}_{\rm BAO}$&1.30&{\color{red}1.88*}&1.45&{\color{red}2.07*} &1.36&{\color{red}2.19*} \\\hline
    $-2\ln\mathcal{L}_{\rm RSD}$&{\color{red}-15.68*}&-17.18&{\color{red}-11.11*} &-17.11&{\color{red}-12.66*} &-15.74\\\hline
    $-2\ln\mathcal{L}_{\rm Lensing}$&{\color{red}-4490.59*} &-4528.23&{\color{red}-4484.87*}&-4527.96&{\color{red}4491.98*} &-4525.14\\\hline
    $-2\ln\mathcal{L}_{\rm SZ}$	&{\color{red}2.18*}&0.00&{\color{red}16.46*} &0.01&{\color{red}12.20*}&0.13\\\hline
    $-2\ln\mathcal{L}$		&9803.96&5261.71&9802.69&5261.50&9802.17&3248.66 \\\hline
  \end{tabular}
  \caption{Marginalised parameter table for the same models presented in Table \ref{table:AST} when the amplitude of the  lensing contribution the CMB temperature power spectrum, $A_{\rm L}$, is allowed to vary. The data combinations are defined in Table~\ref{table:AST} and the numbers in red and denoted * are the values of likelihoods that are not included in the fit calculated to show the tension. The values of  $\sum m_{\nu}$, $m_{\rm sterile}^{\rm eff}$, $\Delta N_{\rm eff}$ are larger than in Table~\ref{table:AST} and $\tau_{\rm R}$ is lower in the final column. }
  \label{table:ASTA}
\end{table}

\subsection{Bayesian evidence}

Another way of quantifying the plausibility of two models, $M_1$ and $M_2$, is Bayesian evidence. This takes into account relative sizes of the model spaces allowed, penalising complicated models with large number of parameters and a significant amount of freedom and favouring simpler models. Prima facie this sounds like a very positive concept. However, in practice it is, as we will see, sensitive to the choice of priors. Typically when a likelihood approach prefers the inclusion of the parameter at $>3\sigma$, the use of Bayesian evidence will come to the same conclusion.

The Bayes factor quantifies the relative plausibility of two models with the same {\it a priori} probability  
  \[\Theta = \frac{P(M_1|d)P(M_2)}{P(M_2|d)P(M_1)},\]
 where $P(M|d)$ is the conditional probability of a model being correct given the data, $d$, and $P(M)$ is the probability of the model being correct~\cite{Jeffreys:1973}. The model probabilities are usually normalised such that $P(M_2)/P(M_1)=1$. When $M_1\subset M_2$ then the Savage-Dickey density ratio can be used to simplify the Bayes factor (see e.g. \cite{Verde:2013} for details) 
 \[\Theta=\frac{P(\psi|d,M_2)}{P(\psi|M_2)}\bigg|_{\psi=\psi_1}\,,\]
 where $\psi$ are the additional parameters in the extended model and $\psi_1$ are their fiducial values in the nested  model.
Therefore, in order to calculate $\Theta$ one only needs the parameter posterior likelihood for the extended model and the probability defined by the prior at the value the parameter would have in the base model. 

The inclusion of active neutrinos can be seen as an addition to $\Lambda$CDM and as such $\psi=\sum m_\nu$ and $M_2$ is $\Lambda$CDM+$\sum m_{\nu}$. Recall that in the vanilla $\Lambda$CDM model $\sum m_\nu$ is set to $0.06{\rm eV}$. The normalised posterior likelihood, $P(\sum m_\nu|d,\Lambda{\rm CDM}+\psi)$ is taken from  the MCMC runs, and we assume a prior range of $\sum m_\nu=[0,3]\,{\rm eV}$. For $d=$ \emph{Planck}+WP+LSSall we find that $\log(\Theta)=-1.8$ implying that a model including active neutrinos is preferred over plain $\Lambda$CDM by odds of around $6:1$. This represents reasonably strong evidence on the Jeffrey's scale~\cite{Jeffreys:1961}. If instead of the prior $[0,3]\,{\rm eV}$, which is not unreasonable but is also not compelling, we use $[0,1]\,{\rm eV}$ or $[0,10]\,{\rm eV}$, we find that $\log(\Theta)= -2.9$ and $-0.6$ respectively. This illustrates the  point about reading too much into the Bayesian evidence as opposed to the likelihood approach: each of the priors is not unreasonable, but the odds of the more complicated model varies from 18:1 to 2:1! The best way to think of the evidence is as another way of quantifying the results of a likelihood analysis rather than as an objective method for distinguishing between models.

\begin{figure}
  \centering
  \includegraphics{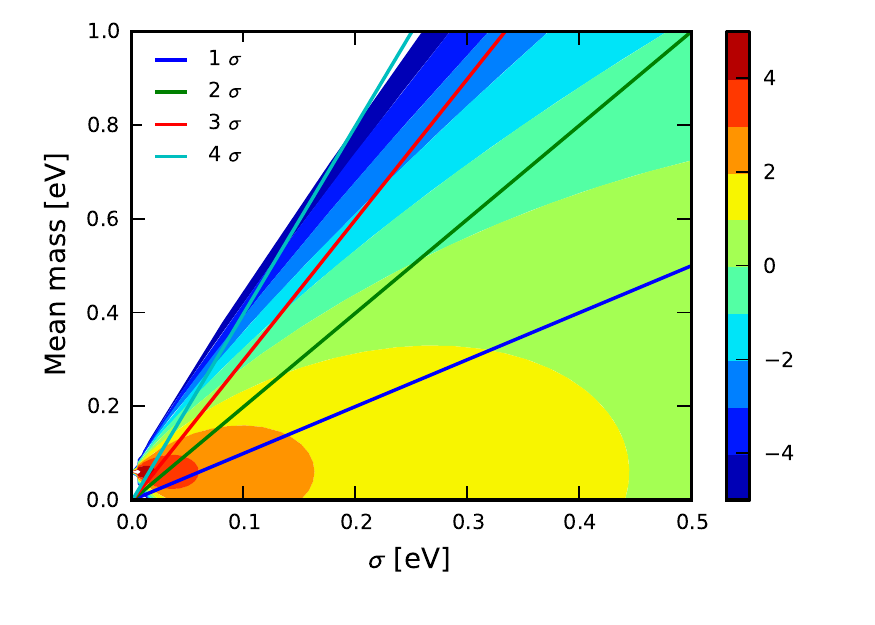}
  \caption{The value of the log of the Bayes factor for a Gaussian probability and fixed flat prior of [0, 3] eV. The 1, 2, 3 and 4 $\sigma$ values are plotted in blue, green, red and light blue respectively.}
  \label{fig:Evidence}
\end{figure}

Fig. \ref{fig:Evidence} shows the value of the log of the Bayes factor when a Gaussian probability is assumed with some flat prior,
\[\log(\Theta)=\log\left(\frac{e^{(x_0-x)^2/2\sigma^2}}{\sigma\sqrt{2\pi}}\times\frac{1}{P_{\rm Prior}}\right),\]
where $\sigma$ is the standard deviation, $x$ is the mean mass, $x_0$ is the fiducial mass and $P_{\rm Prior}$ is the prior range normalised to 1. For $\Lambda{\rm CDM}+\sum m_\nu$ then $x=0.06{\rm eV}$ and with the prior $[0,3]\,{\rm eV}$, $P_{\rm Prior}=1/3$. One can see that, for $\sum m_\nu=0.357\pm0.099$ obtained from \emph{Planck}+WP+LSSall, the value of the Bayes factor is $\ln(\Theta)\sim-2$ and has between 3-4$\sigma$ significance (red and light blue lines). As such, not only are the Bayesian statistics showing reasonably strong odds, but so is the statistical likelihood. If the prior range is changed to some larger value, the Bayes factor becomes larger and starts supporting no change from standard $\Lambda{\rm CDM}$, but the statistical likelihood would continue to show that active neutrinos are significant within the 3-4$\sigma$ range. 
\medskip

The sterile neutrino case is slightly more complicated  as there are two extra parameters, so that $\psi=(m_{\rm eff}^{\rm sterile},\Delta N_{\rm eff}$). The 2D posterior likelihood, $P(\psi|d,\Lambda{\rm CDM}+\psi)$ is again obtained from MCMC with the $\Lambda$CDM parameter values being $m_{\rm eff}^{\rm sterile}=0\,{\rm eV}$ and $N_{\rm eff}=3.046$. The priors used are $m_{\rm eff}^{\rm sterile}=[0,3]\,{\rm eV}$ and $N_{\rm eff}=[3.046,10]$. Again using $d=$ \emph{Planck}+WP+LSSall then $\log(\Theta)=-2.67$, which strongly supports the addition of sterile neutrinos over $\Lambda$CDM. This is in contrast to the values presented in \cite{Boris}. For a similar data combination they find that $\log(\Theta)\approx 1$. Since they use a similar prior range, $m_{\rm eff}^{\rm sterile}=[0,3]\,{\rm eV}$, it appears that the discrepancy is due to the posterior likelihood for $m_{\rm eff}^{\rm sterile}$. Although no specific numbers are presented, it is clear from the right-hand panel of Fig.~1 in \cite{Boris} that their constraint is much weaker than the $m_{\rm eff}^{\rm sterile}=0.67\pm 0.18$ we report here. We note that our numbers are compatible with those presented in \cite{Battye:2013,Wyman:2013,Hamann:2013} all of which suggest $\sim 4\sigma$ preferences for sterile neutrinos, albeit for slightly different data combinations.

\subsection{Active neutrino mass hierarchy}

Cosmological limits on $\sum m_{\nu}$ for active neutrinos are important since they can be combined with square differences between the neutrino masses obtained from atmospheric, $\Delta m_{\rm A}^2$, and solar neutrino measurements, $\Delta m_{\rm S}^2$, in order to calculate the masses of the individual eigenstates, $(m_{\nu_1},m_{\nu_2},m_{\nu_3})$. An important question to answer is whether the masses have a ``normal'', ``inverted'' or ``degenerate'' hierarchy. From \cite{Capozzi:2013csa}  the difference between $m_{\nu_3}^2$ and $m_{\nu_1}^2$ is given by $\Delta m_{\rm A}^2=2.4_{-0.1}^{+0.1}\times10^{-3}$eV$^{2}$ (i.e. an uncertainty of $\sim$ 4\%) and the difference between $m_{\nu_2}^2$ and $m_{\nu_1}^2$ is given by $\Delta m_{\rm S}^2=7.5_{-0.2}^{+0.3}\times10^{-5}$eV$^2$ (i.e. an uncertainty of $\sim$ 3\%). For normal hierarchy, $m_{\nu_1}^2$ and $m_{\nu_2}^2$ are much less than $m_{\nu_3}^2$ and the mass eigenstates are given by $m_{\nu_2}^2=m_{\nu_1}^2+\Delta m_{\rm S}^2$ and $m_{\nu_3}^2=m_{\nu_1}^2+\Delta m_{\rm A}^2$ with $m_{\nu_1}$ being the lowest mass eigenstate, and in the inverted hierarchy, $m_{\nu_3}^2$ is the lowest mass eigenstate with a mass much less than $m_{\nu_1}^2=m_{\nu_3}^2+\Delta m_{\rm A}^2-\Delta m_{\rm S}^2$ and $m_{\nu_2}^2=m_{\nu_3}^2+\Delta m_{\rm A}^2$.

\begin{figure}
\centering
\includegraphics{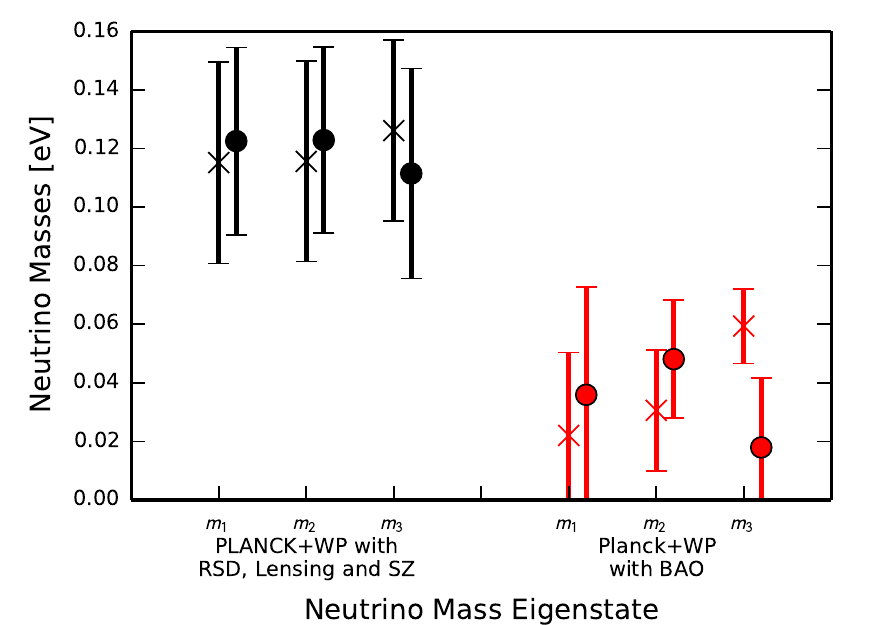}
\caption{Masses of the neutrino eigenstates using \emph{Planck}+WP+BAO (red) and \emph{Planck}+WP+LSSall (black), together with results from oscillation experiments~\cite{Capozzi:2013csa}. These allow for two solutions, called normal (crosses) and inverted (circles) hierarchies. When the higher $\sum m_\nu$ value from \emph{Planck}+WP+LSSall is used then the eigenstates become degenerate.}
\label{fig:masses}
\end{figure}

In Fig.~\ref{fig:masses} the individual neutrino masses are shown for $\sum m_\nu$ given the limits from \emph{Planck}+WP+BAO and \emph{Planck}+WP+LSSall, i.e. the two collections of data used most often throughout the paper. The normal hierarchy equations are plotted with crosses and the inverted with circles. \emph{Planck}+WP+BAO predicts the mass of three active neutrinos to be $<0.258\,{\rm eV}$ with $m_{\nu_3}$ greater than $m_{\nu_1}$ and $m_{\nu_2}$ indicating that, if this limit is correct, then a ``normal'' hierarchy is preferred, although ``inverted'' or ``degenerate'' hierarchy are by no means excluded. The preferred masses are $m_{\nu_1}=(0.022\pm0.028)\,{\rm eV}$, $m_{\nu_2}=(0.031\pm0.021)\,{\rm eV}$ and $m_{\nu_3}=(0.059\pm0.013)\,{\rm eV}$.  The same is true when looking at the inverted hierarchy equation, where clear preference is seen for ``inverted'' whilst ``normal'' and ``degenerate'' hierarchies are not ruled out. The masses in this case are $m_{\nu_1}=(0.036\pm0.028)\,{\rm eV}$, $m_{\nu_2}=(0.048\pm0.020)\,{\rm eV}$ and $m_{\nu_3}=(0.018\pm0.024)\,{\rm eV}$. Interestingly, when looking at the inverted hierarchy equation, $m_{\nu_1}$ is unbounded from below like $m_{\nu_3}$ is whereas $m_{\nu_3}$ is statistically unlikely to have zero mass in the normal hierarchy case, perhaps suggesting slightly more preference for normal hierarchy. \emph{Planck}+WP+LSSall predicts $\sum m_{\nu}=(0.357\pm0.099){\rm,eV}$ which is more consistent with a degenerate hierarchy for both the normal and inverted equations. The preferred masses in this case are $m_{\nu_1}=(0.115\pm0.034)\,{\rm eV}$ or $m_{\nu_1}=(0.123\pm0.032)\,{\rm eV}$, $m_{\nu_2}=(0.116\pm0.034)\,{\rm eV}$ or $m_{\nu_2}=(0.123\pm0.032)\,{\rm eV}$ and $m_{\nu_3}=(0.126\pm0.031)\,{\rm eV}$ or $m_{\nu_3}=(0.111\pm0.036)\,{\rm eV}$ for normal and inverted equations respectively. Each indicate more than $3\sigma$ preference for non-zero neutrino mass. We note, however, there is strong correlation between the probability distribution for each eigenstate. In both the normal and inverted hierarchy cases the masses appear to be more or less degenerate. We are unable to say, with any certainty, which hierarchy is preferred from our results, the constraining power of which is most likely to come from neutrino oscillation experiments, but in general, the higher the summed neutrino mass the more degenerate each mass eigenstate becomes. 
 
\section{Conclusions} \label{sec:conclusions}

We have found a tension in excess of  $5\sigma$ between the observations of \emph{Planck}  and lensing, SZ cluster and RSD data. Moreover, this  is:
\begin{enumerate}
\item still significant when using WMAP+highL instead of {\it Planck};
\item still significant when excluding SZ cluster counts and only using lensing and RSD data. 
\end{enumerate}

An obvious candidate to alleviate this tension are massive neutrinos, either active or sterile.  We find that the addition of three active neutrinos, with a combined mass of $\sum m_\nu = 0.357\pm0.099{\rm eV}$, or a sterile neutrino with $m_{\rm sterile}^{\rm eff}= 0.66\pm 0.18{\rm eV}$ and $\Delta N_{\rm eff}=0.32\pm0.21$, helps reduce the discrepancy to the $\sim 2\sigma$ level. The residual tension is at the expense of a degraded fit to {\it Planck} data, which doesn't favour such neutrino masses. Therefore, although there is significant preference for modifications to the neutrino sector over $\Lambda$CDM, the solution isn't perfect. 

We have also discussed alternatives, such as modifications to the reionization history and primordial power spectrum. We find that if one ignores WMAP polarization, LSS prefers a lower value of $\tau_{\rm R}=0.049\pm 0.021$. This might be attractive astrophysically as it agrees with the end of reionization from the discovery of a complete Gunn-Peterson trough at $z=6.28$. However, this model also suffers problems: one would need to understand any systematics in the WMAP polarization measurements which would suggest higher $\tau_{\rm R}$, and the {\it Planck} temperature data is also somewhat inconsistent with lower values of $\tau_{\rm R}$. We have found that a modified power spectrum model can also reduce small scale power, and indeed the global fit to {\em Planck}+LSS data is better than the neutrino model. However, this model is somewhat ad-hoc and more importantly, seems to be excluded with the addition of high $\ell$ data. 

Looking forward, one can be optimistic that the origin of this tension will be resolved. {\it Planck} will soon release new data, including polarisation, which will fix the model  preferred by the CMB to even higher accuracy. {\it Planck} will also provide improved lensing reconstruction maps,  covering a larger $\ell$ range. It will also provide a larger catalogue of SZ clusters with improved determination of the mass bias. For RSD's we can soon  look forward to the extended Baryon Oscillation Spectroscopic Survey (eBOSS) (see e.g.~\cite{Weinberg:2013raj} for an overview), which will measure $f \sigma_8$ with an accuracy of $\lesssim 3.5 \%$ in several redshift bins to $z=2.2$. The Hobby-Eberly Telescope Dark Energy Experiment (HETDEX) will also measure $f \sigma_8$ at high redshift, which an accuracy of  $\lesssim 2 \%$ at $z=2.3$. Both these experiments are expected to produce results within a few years. The Stage III (Dark Energy Task Force~\cite{Albrecht:2006um}) galaxy weak lensing surveys are now underway and we can look forward to their first lensing results within the next year. The Kilo-Degree Survey (KiDS) has so far imaged about on fifth of its total of 1500 square degrees. The Dark Energy Survey (DES) has imaged about one fifth of its total of 5000 square degrees. The Subaru Hyper-Suprime Camera (HSC) has began it's coverage of 1000 square degrees to significantly greater depth. These will provide improved, independent results from CFHTLenS, in terms of the area on the sky and the shear pipelines. If, after this, tensions with the CMB {\em still} remain, we may look forward to the tantalising prospect of new physics. We have shown that modifications to the neutrino sector do not provide a perfect solution to the problem -- other more exotic possibilities might include modifications to the dark sector of the Universe.

\medskip

{\it Acknowledgements:} ~This research was supported by STFC. We thank Florian Beutler, Sarah Bridle, Boris Leistedt, Hiranya Peiris and Licia Verde for helpful discussions.  We acknowledge the use of the {\tt CAMB}~\cite{2000ApJ...538..473L} and {\tt COSMOMC}~\cite{2002PhRvD..66j3511L} codes, and  the use of {\it Planck} data. The development of {\it Planck} has been supported by: ESA; CNES and 
CNRS/INSU-IN2P3-INP (France); ASI, CNR, and INAF (Italy); NASA and DoE 
(USA); STFC and UKSA (UK); CSIC, MICINN, JA and RES (Spain); Tekes, AoF 
and CSC (Finland); DLR and MPG (Germany); CSA (Canada); DTU Space 
(Denmark); SER/SSO (Switzerland); RCN (Norway); SFI (Ireland); FCT/MCTES 
(Portugal); and PRACE (EU). A description of the {\it Planck} Collaboration 
and a list of its members, including the technical or scientific 
activities in which they have been involved, can be found at 
\url{http://www.sciops.esa.int/index.php?project=planck&page=Planck_Collaboration} 

\bibliography{bib}

\def\eprinttmppp@#1arXiv:@{#1}
\providecommand{\arxivlink[1]}{\href{http://arxiv.org/abs/#1}{arXiv:#1}}
\def\eprinttmp@#1arXiv:#2 [#3]#4@{\ifthenelse{\equal{#3}{x}}{\ifthenelse{
\equal{#1}{}}{\arxivlink{\eprinttmppp@#2@}}{\arxivlink{#1}}}{\arxivlink{#2}
  [#3]}}
\providecommand{\eprintlink}[1]{\eprinttmp@#1arXiv: [x]@}
\renewcommand{\eprint}[1]{\eprintlink{#1}}
\providecommand{\adsurl}[1]{\href{#1}{ADS}}
\renewcommand{\bibinfo}[2]{\ifthenelse{\equal{#1}{isbn}}{\href{http://cosmologist.info/ISBN/#2}{#2}}{#2}}
\begin{thebibliography}{84}
\expandafter\ifx\csname natexlab\endcsname\relax\def\natexlab#1{#1}\fi
\expandafter\ifx\csname bibnamefont\endcsname\relax
  \def\bibnamefont#1{#1}\fi
\expandafter\ifx\csname bibfnamefont\endcsname\relax
  \def\bibfnamefont#1{#1}\fi
\expandafter\ifx\csname citenamefont\endcsname\relax
  \def\citenamefont#1{#1}\fi
\expandafter\ifx\csname url\endcsname\relax
  \def\url#1{\texttt{#1}}\fi
\expandafter\ifx\csname urlprefix\endcsname\relax\def\urlprefix{URL }\fi

\bibitem[{\citenamefont{{Riess} et~al.}(1998)\citenamefont{{Riess},
  {Filippenko}, {Challis}, {Clocchiatti}, {Diercks}, {Garnavich}, {Gilliland},
  {Hogan}, {Jha}, {Kirshner} et~al.}}]{1998AJ....116.1009R}
\bibinfo{author}{\bibfnamefont{A.~G.} \bibnamefont{{Riess}}},
  \bibinfo{author}{\bibfnamefont{A.~V.} \bibnamefont{{Filippenko}}},
  \bibinfo{author}{\bibfnamefont{P.}~\bibnamefont{{Challis}}},
  \bibinfo{author}{\bibfnamefont{A.}~\bibnamefont{{Clocchiatti}}},
  \bibnamefont{et~al.}, \bibinfo{journal}{Astronomical Journal}
  \textbf{\bibinfo{volume}{116}}, \bibinfo{pages}{1009} (\bibinfo{year}{1998}),
  \eprint{astro-ph/9805201}.

\bibitem[{\citenamefont{{Perlmutter} et~al.}(1999)\citenamefont{{Perlmutter},
  {Aldering}, {Goldhaber}, {Knop}, {Nugent}, {Castro}, {Deustua}, {Fabbro},
  {Goobar}, {Groom} et~al.}}]{1999ApJ...517..565P}
\bibinfo{author}{\bibfnamefont{S.}~\bibnamefont{{Perlmutter}}},
  \bibinfo{author}{\bibfnamefont{G.}~\bibnamefont{{Aldering}}},
  \bibinfo{author}{\bibfnamefont{G.}~\bibnamefont{{Goldhaber}}},
  \bibinfo{author}{\bibfnamefont{R.~A.} \bibnamefont{{Knop}}},
  \bibnamefont{et~al.}, \bibinfo{journal}{\apj} \textbf{\bibinfo{volume}{517}},
  \bibinfo{pages}{565} (\bibinfo{year}{1999}), \eprint{astro-ph/9812133}.

\bibitem[{\citenamefont{{Henry}}(1997)}]{1997ApJ...489L...1H}
\bibinfo{author}{\bibfnamefont{J.~P.} \bibnamefont{{Henry}}},
  \bibinfo{journal}{Astrophysical Journal Letters}
  \textbf{\bibinfo{volume}{489}}, \bibinfo{pages}{L1} (\bibinfo{year}{1997}),
  \adsurl{http://ukads.nottingham.ac.uk/abs/1997ApJ...489L...1H}.

\bibitem[{\citenamefont{{Eke} et~al.}(1998)\citenamefont{{Eke}, {Cole},
  {Frenk}, and {Patrick Henry}}}]{1998MNRAS.298.1145E}
\bibinfo{author}{\bibfnamefont{V.~R.} \bibnamefont{{Eke}}},
  \bibinfo{author}{\bibfnamefont{S.}~\bibnamefont{{Cole}}},
  \bibinfo{author}{\bibfnamefont{C.~S.} \bibnamefont{{Frenk}}},
  \bibnamefont{and} \bibinfo{author}{\bibfnamefont{J.}~\bibnamefont{{Patrick
  Henry}}}, \bibinfo{journal}{\mnras} \textbf{\bibinfo{volume}{298}},
  \bibinfo{pages}{1145} (\bibinfo{year}{1998}), \eprint{astro-ph/9802350}.

\bibitem[{\citenamefont{{Efstathiou} et~al.}(1990)\citenamefont{{Efstathiou},
  {Sutherland}, and {Maddox}}}]{1990Natur.348..705E}
\bibinfo{author}{\bibfnamefont{G.}~\bibnamefont{{Efstathiou}}},
  \bibinfo{author}{\bibfnamefont{W.~J.} \bibnamefont{{Sutherland}}},
  \bibnamefont{and} \bibinfo{author}{\bibfnamefont{S.~J.}
  \bibnamefont{{Maddox}}}, \bibinfo{journal}{\nat}
  \textbf{\bibinfo{volume}{348}}, \bibinfo{pages}{705} (\bibinfo{year}{1990}),
  \adsurl{http://ukads.nottingham.ac.uk/abs/1990Natur.348..705E}.

\bibitem[{\citenamefont{{Bridle} et~al.}(1999)\citenamefont{{Bridle}, {Eke},
  {Lahav}, {Lasenby}, {Hobson}, {Cole}, {Frenk}, and
  {Henry}}}]{1999MNRAS.310..565B}
\bibinfo{author}{\bibfnamefont{S.~L.} \bibnamefont{{Bridle}}},
  \bibinfo{author}{\bibfnamefont{V.~R.} \bibnamefont{{Eke}}},
  \bibinfo{author}{\bibfnamefont{O.}~\bibnamefont{{Lahav}}},
  \bibinfo{author}{\bibfnamefont{A.~N.} \bibnamefont{{Lasenby}}},
  \bibnamefont{et~al.}, \bibinfo{journal}{\mnras}
  \textbf{\bibinfo{volume}{310}}, \bibinfo{pages}{565} (\bibinfo{year}{1999}),
  \eprint{astro-ph/9903472}.

\bibitem[{\citenamefont{{Planck Collaboration}
  et~al.}(2013{\natexlab{a}})\citenamefont{{Planck Collaboration}, {Ade},
  {Aghanim}, {Armitage-Caplan}, {Arnaud}, {Ashdown}, {Atrio-Barandela},
  {Aumont}, {Baccigalupi}, {Banday} et~al.}}]{2013arXiv1303.5062P}
\bibinfo{author}{\bibnamefont{{Planck Collaboration}}},
  \bibinfo{author}{\bibfnamefont{P.~A.~R.} \bibnamefont{{Ade}}},
  \bibinfo{author}{\bibfnamefont{N.}~\bibnamefont{{Aghanim}}},
  \bibinfo{author}{\bibfnamefont{C.}~\bibnamefont{{Armitage-Caplan}}},
  \bibnamefont{et~al.}, \bibinfo{journal}{ArXiv e-prints}
  (\bibinfo{year}{2013}{\natexlab{a}}), \eprint{1303.5062}.

\bibitem[{\citenamefont{{Bennett} et~al.}(2013)\citenamefont{{Bennett},
  {Larson}, {Weiland}, {Jarosik}, {Hinshaw}, {Odegard}, {Smith}, {Hill},
  {Gold}, {Halpern} et~al.}}]{2013ApJS..208...20B}
\bibinfo{author}{\bibfnamefont{C.~L.} \bibnamefont{{Bennett}}},
  \bibinfo{author}{\bibfnamefont{D.}~\bibnamefont{{Larson}}},
  \bibinfo{author}{\bibfnamefont{J.~L.} \bibnamefont{{Weiland}}},
  \bibinfo{author}{\bibfnamefont{N.}~\bibnamefont{{Jarosik}}},
  \bibnamefont{et~al.}, \bibinfo{journal}{Astrophysical Journal Supp.}
  \textbf{\bibinfo{volume}{208}}, \bibinfo{eid}{20} (\bibinfo{year}{2013}),
  \eprint{1212.5225}.

\bibitem[{\citenamefont{Das et~al.}(2013)\citenamefont{Das, Louis, Nolta,
  Addison, Battistelli et~al.}}]{Das:2013zf}
\bibinfo{author}{\bibfnamefont{S.}~\bibnamefont{Das}},
  \bibinfo{author}{\bibfnamefont{T.}~\bibnamefont{Louis}},
  \bibinfo{author}{\bibfnamefont{M.~R.} \bibnamefont{Nolta}},
  \bibinfo{author}{\bibfnamefont{G.~E.} \bibnamefont{Addison}},
  \bibnamefont{et~al.} (\bibinfo{year}{2013}), \eprint{1301.1037}.

\bibitem[{\citenamefont{{Reichardt} et~al.}(2012)\citenamefont{{Reichardt},
  {Shaw}, {Zahn}, {Aird}, {Benson}, {Bleem}, {Carlstrom}, {Chang}, {Cho},
  {Crawford} et~al.}}]{2012ApJ...755...70R}
\bibinfo{author}{\bibfnamefont{C.~L.} \bibnamefont{{Reichardt}}},
  \bibinfo{author}{\bibfnamefont{L.}~\bibnamefont{{Shaw}}},
  \bibinfo{author}{\bibfnamefont{O.}~\bibnamefont{{Zahn}}},
  \bibinfo{author}{\bibfnamefont{K.~A.} \bibnamefont{{Aird}}},
  \bibnamefont{et~al.}, \bibinfo{journal}{\apj} \textbf{\bibinfo{volume}{755}},
  \bibinfo{eid}{70} (\bibinfo{year}{2012}), \eprint{1111.0932}.

\bibitem[{\citenamefont{Battye and Moss}(2014)}]{Battye:2013}
\bibinfo{author}{\bibfnamefont{R.~A.} \bibnamefont{Battye}} \bibnamefont{and}
  \bibinfo{author}{\bibfnamefont{A.}~\bibnamefont{Moss}},
  \bibinfo{journal}{Phys.Rev.Lett.} \textbf{\bibinfo{volume}{112}},
  \bibinfo{pages}{051303} (\bibinfo{year}{2014}), \eprint{1308.5870}.

\bibitem[{\citenamefont{Wyman et~al.}(2014)\citenamefont{Wyman, Rudd,
  Vanderveld, and Hu}}]{Wyman:2013}
\bibinfo{author}{\bibfnamefont{M.}~\bibnamefont{Wyman}},
  \bibinfo{author}{\bibfnamefont{D.~H.} \bibnamefont{Rudd}},
  \bibinfo{author}{\bibfnamefont{R.~A.} \bibnamefont{Vanderveld}},
  \bibnamefont{and} \bibinfo{author}{\bibfnamefont{W.}~\bibnamefont{Hu}},
  \bibinfo{journal}{Phys.Rev.Lett.} \textbf{\bibinfo{volume}{112}},
  \bibinfo{pages}{051302} (\bibinfo{year}{2014}), \eprint{1307.7715}.

\bibitem[{\citenamefont{Hamann and Hasenkamp}(2013)}]{Hamann:2013}
\bibinfo{author}{\bibfnamefont{J.}~\bibnamefont{Hamann}} \bibnamefont{and}
  \bibinfo{author}{\bibfnamefont{J.}~\bibnamefont{Hasenkamp}},
  \bibinfo{journal}{JCAP} \textbf{\bibinfo{volume}{1310}}, \bibinfo{pages}{044}
  (\bibinfo{year}{2013}), \eprint{1308.3255}.

\bibitem[{\citenamefont{{Planck Collaboration}
  et~al.}(2013{\natexlab{b}})\citenamefont{{Planck Collaboration}, {Ade},
  {Aghanim}, {Armitage-Caplan}, {Arnaud}, {Ashdown}, {Atrio-Barandela},
  {Aumont}, {Baccigalupi}, {Banday} et~al.}}]{2013arXiv1303.5080P}
\bibinfo{author}{\bibnamefont{{Planck Collaboration}}},
  \bibinfo{author}{\bibfnamefont{P.~A.~R.} \bibnamefont{{Ade}}},
  \bibinfo{author}{\bibfnamefont{N.}~\bibnamefont{{Aghanim}}},
  \bibinfo{author}{\bibfnamefont{C.}~\bibnamefont{{Armitage-Caplan}}},
  \bibnamefont{et~al.}, \bibinfo{journal}{ArXiv e-prints}
  (\bibinfo{year}{2013}{\natexlab{b}}), \eprint{1303.5080}.

\bibitem[{\citenamefont{Riess et~al.}(2011{\natexlab{a}})\citenamefont{Riess,
  Macri, Casertano, Lampeitl, Ferguson et~al.}}]{Riess:2011yx}
\bibinfo{author}{\bibfnamefont{A.~G.} \bibnamefont{Riess}},
  \bibinfo{author}{\bibfnamefont{L.}~\bibnamefont{Macri}},
  \bibinfo{author}{\bibfnamefont{S.}~\bibnamefont{Casertano}},
  \bibinfo{author}{\bibfnamefont{H.}~\bibnamefont{Lampeitl}},
  \bibnamefont{et~al.}, \bibinfo{journal}{Astrophys.J.}
  \textbf{\bibinfo{volume}{730}}, \bibinfo{pages}{119}
  (\bibinfo{year}{2011}{\natexlab{a}}), \eprint{1103.2976}.

\bibitem[{\citenamefont{{Planck Collaboration}
  et~al.}(2013{\natexlab{c}})\citenamefont{{Planck Collaboration}, {Ade},
  {Aghanim}, {Armitage-Caplan}, {Arnaud}, {Ashdown}, {Atrio-Barandela},
  {Aumont}, {Baccigalupi}, {Banday} et~al.}}]{2013arXiv1303.5076P}
\bibinfo{author}{\bibnamefont{{Planck Collaboration}}},
  \bibinfo{author}{\bibfnamefont{P.~A.~R.} \bibnamefont{{Ade}}},
  \bibinfo{author}{\bibfnamefont{N.}~\bibnamefont{{Aghanim}}},
  \bibinfo{author}{\bibfnamefont{C.}~\bibnamefont{{Armitage-Caplan}}},
  \bibnamefont{et~al.}, \bibinfo{journal}{ArXiv e-prints}
  (\bibinfo{year}{2013}{\natexlab{c}}), \eprint{1303.5076}.

\bibitem[{\citenamefont{{Planck Collaboration}
  et~al.}(2013{\natexlab{d}})\citenamefont{{Planck Collaboration}, {Ade},
  {Aghanim}, {Armitage-Caplan}, {Arnaud}, {Ashdown}, {Atrio-Barandela},
  {Aumont}, {Baccigalupi}, {Banday} et~al.}}]{2013arXiv1303.5075P}
\bibinfo{author}{\bibnamefont{{Planck Collaboration}}},
  \bibinfo{author}{\bibfnamefont{P.~A.~R.} \bibnamefont{{Ade}}},
  \bibinfo{author}{\bibfnamefont{N.}~\bibnamefont{{Aghanim}}},
  \bibinfo{author}{\bibfnamefont{C.}~\bibnamefont{{Armitage-Caplan}}},
  \bibnamefont{et~al.}, \bibinfo{journal}{ArXiv e-prints}
  (\bibinfo{year}{2013}{\natexlab{d}}), \eprint{1303.5075}.

\bibitem[{\citenamefont{{Hinshaw} et~al.}(2013)\citenamefont{{Hinshaw},
  {Larson}, {Komatsu}, {Spergel}, {Bennett}, {Dunkley}, {Nolta}, {Halpern},
  {Hill}, {Odegard} et~al.}}]{2013ApJS..208...19H}
\bibinfo{author}{\bibfnamefont{G.}~\bibnamefont{{Hinshaw}}},
  \bibinfo{author}{\bibfnamefont{D.}~\bibnamefont{{Larson}}},
  \bibinfo{author}{\bibfnamefont{E.}~\bibnamefont{{Komatsu}}},
  \bibinfo{author}{\bibfnamefont{D.~N.} \bibnamefont{{Spergel}}},
  \bibnamefont{et~al.}, \bibinfo{journal}{Astrophysical Journal Supp.}
  \textbf{\bibinfo{volume}{208}}, \bibinfo{eid}{19} (\bibinfo{year}{2013}),
  \eprint{1212.5226}.

\bibitem[{\citenamefont{Jones et~al.}(2004)\citenamefont{Jones, Saunders,
  Colless, Read, Parker et~al.}}]{Jones:2004zy}
\bibinfo{author}{\bibfnamefont{D.~H.} \bibnamefont{Jones}},
  \bibinfo{author}{\bibfnamefont{W.}~\bibnamefont{Saunders}},
  \bibinfo{author}{\bibfnamefont{M.}~\bibnamefont{Colless}},
  \bibinfo{author}{\bibfnamefont{M.~A.} \bibnamefont{Read}},
  \bibnamefont{et~al.}, \bibinfo{journal}{Mon.Not.Roy.Astron.Soc.}
  \textbf{\bibinfo{volume}{355}}, \bibinfo{pages}{747} (\bibinfo{year}{2004}),
  \eprint{astro-ph/0403501}.

\bibitem[{\citenamefont{Beutler et~al.}(2011)\citenamefont{Beutler, Blake,
  Colless, Jones, Staveley-Smith et~al.}}]{Beutler:2011hx}
\bibinfo{author}{\bibfnamefont{F.}~\bibnamefont{Beutler}},
  \bibinfo{author}{\bibfnamefont{C.}~\bibnamefont{Blake}},
  \bibinfo{author}{\bibfnamefont{M.}~\bibnamefont{Colless}},
  \bibinfo{author}{\bibfnamefont{D.~H.} \bibnamefont{Jones}},
  \bibnamefont{et~al.}, \bibinfo{journal}{Mon.Not.Roy.Astron.Soc.}
  \textbf{\bibinfo{volume}{416}}, \bibinfo{pages}{3017} (\bibinfo{year}{2011}),
  \eprint{1106.3366}.

\bibitem[{\citenamefont{Padmanabhan et~al.}(2012)\citenamefont{Padmanabhan, Xu,
  Eisenstein, Scalzo, Cuesta et~al.}}]{Padmanabhan:2012hf}
\bibinfo{author}{\bibfnamefont{N.}~\bibnamefont{Padmanabhan}},
  \bibinfo{author}{\bibfnamefont{X.}~\bibnamefont{Xu}},
  \bibinfo{author}{\bibfnamefont{D.~J.} \bibnamefont{Eisenstein}},
  \bibinfo{author}{\bibfnamefont{R.}~\bibnamefont{Scalzo}},
  \bibnamefont{et~al.}, \bibinfo{journal}{Mon.Not.Roy.Astron.Soc.}
  \textbf{\bibinfo{volume}{427}}, \bibinfo{pages}{2132} (\bibinfo{year}{2012}),
  \eprint{1202.0090}.

\bibitem[{\citenamefont{Anderson et~al.}(2013)\citenamefont{Anderson, Aubourg,
  Bailey, Bizyaev, Blanton et~al.}}]{Anderson:2012sa}
\bibinfo{author}{\bibfnamefont{L.}~\bibnamefont{Anderson}},
  \bibinfo{author}{\bibfnamefont{E.}~\bibnamefont{Aubourg}},
  \bibinfo{author}{\bibfnamefont{S.}~\bibnamefont{Bailey}},
  \bibinfo{author}{\bibfnamefont{D.}~\bibnamefont{Bizyaev}},
  \bibnamefont{et~al.}, \bibinfo{journal}{Mon.Not.Roy.Astron.Soc.}
  \textbf{\bibinfo{volume}{427}}, \bibinfo{pages}{3435} (\bibinfo{year}{2013}),
  \eprint{1203.6594}.

\bibitem[{\citenamefont{Saro et~al.}(2013)}]{Saro:2013}
\bibinfo{author}{\bibfnamefont{A.}~\bibnamefont{Saro}} \bibnamefont{et~al.}
  (\bibinfo{collaboration}{SPT Collaboration}) (\bibinfo{year}{2013}),
  \eprint{1312.2462}.

\bibitem[{\citenamefont{Hasselfield et~al.}(2013)\citenamefont{Hasselfield,
  Hilton, Marriage, Addison, Barrientos et~al.}}]{Hasselfield:2013}
\bibinfo{author}{\bibfnamefont{M.}~\bibnamefont{Hasselfield}},
  \bibinfo{author}{\bibfnamefont{M.}~\bibnamefont{Hilton}},
  \bibinfo{author}{\bibfnamefont{T.~A.} \bibnamefont{Marriage}},
  \bibinfo{author}{\bibfnamefont{G.~E.} \bibnamefont{Addison}},
  \bibnamefont{et~al.}, \bibinfo{journal}{JCAP}
  \textbf{\bibinfo{volume}{1307}}, \bibinfo{pages}{008} (\bibinfo{year}{2013}),
  \eprint{1301.0816}.

\bibitem[{\citenamefont{{Vikhlinin} et~al.}(2009)\citenamefont{{Vikhlinin},
  {Kravtsov}, {Burenin}, {Ebeling}, {Forman}, {Hornstrup}, {Jones}, {Murray},
  {Nagai}, {Quintana} et~al.}}]{vik09}
\bibinfo{author}{\bibfnamefont{A.}~\bibnamefont{{Vikhlinin}}},
  \bibinfo{author}{\bibfnamefont{A.~V.} \bibnamefont{{Kravtsov}}},
  \bibinfo{author}{\bibfnamefont{R.~A.} \bibnamefont{{Burenin}}},
  \bibinfo{author}{\bibfnamefont{H.}~\bibnamefont{{Ebeling}}},
  \bibnamefont{et~al.}, \bibinfo{journal}{\apj} \textbf{\bibinfo{volume}{692}},
  \bibinfo{pages}{1060} (\bibinfo{year}{2009}), \eprint{0812.2720}.

\bibitem[{\citenamefont{{Rozo} et~al.}(2010)\citenamefont{{Rozo}, {Wechsler},
  {Rykoff}, {Annis}, {Becker}, {Evrard}, {Frieman}, {Hansen}, {Hao}, {Johnston}
  et~al.}}]{2010ApJ...708..645R}
\bibinfo{author}{\bibfnamefont{E.}~\bibnamefont{{Rozo}}},
  \bibinfo{author}{\bibfnamefont{R.~H.} \bibnamefont{{Wechsler}}},
  \bibinfo{author}{\bibfnamefont{E.~S.} \bibnamefont{{Rykoff}}},
  \bibinfo{author}{\bibfnamefont{J.~T.} \bibnamefont{{Annis}}},
  \bibnamefont{et~al.}, \bibinfo{journal}{\apj} \textbf{\bibinfo{volume}{708}},
  \bibinfo{pages}{645} (\bibinfo{year}{2010}), \eprint{0902.3702}.

\bibitem[{\citenamefont{{Heymans} et~al.}(2012)\citenamefont{{Heymans}, {Van
  Waerbeke}, {Miller}, {Erben}, {Hildebrandt}, {Hoekstra}, {Kitching},
  {Mellier}, {Simon}, {Bonnett} et~al.}}]{2012MNRAS.427..146H}
\bibinfo{author}{\bibfnamefont{C.}~\bibnamefont{{Heymans}}},
  \bibinfo{author}{\bibfnamefont{L.}~\bibnamefont{{Van Waerbeke}}},
  \bibinfo{author}{\bibfnamefont{L.}~\bibnamefont{{Miller}}},
  \bibinfo{author}{\bibfnamefont{T.}~\bibnamefont{{Erben}}},
  \bibnamefont{et~al.}, \bibinfo{journal}{\mnras}
  \textbf{\bibinfo{volume}{427}}, \bibinfo{pages}{146} (\bibinfo{year}{2012}),
  \eprint{1210.0032}.

\bibitem[{\citenamefont{Heymans et~al.}(2013)\citenamefont{Heymans, Grocutt,
  Heavens, Kilbinger, Kitching et~al.}}]{Heymans:2013fya}
\bibinfo{author}{\bibfnamefont{C.}~\bibnamefont{Heymans}},
  \bibinfo{author}{\bibfnamefont{E.}~\bibnamefont{Grocutt}},
  \bibinfo{author}{\bibfnamefont{A.}~\bibnamefont{Heavens}},
  \bibinfo{author}{\bibfnamefont{M.}~\bibnamefont{Kilbinger}},
  \bibnamefont{et~al.} (\bibinfo{year}{2013}), \eprint{1303.1808}.

\bibitem[{\citenamefont{{Kilbinger} et~al.}(2013)\citenamefont{{Kilbinger},
  {Fu}, {Heymans}, {Simpson}, {Benjamin}, {Erben}, {Harnois-D{\'e}raps},
  {Hoekstra}, {Hildebrandt}, {Kitching} et~al.}}]{2013MNRAS.430.2200K}
\bibinfo{author}{\bibfnamefont{M.}~\bibnamefont{{Kilbinger}}},
  \bibinfo{author}{\bibfnamefont{L.}~\bibnamefont{{Fu}}},
  \bibinfo{author}{\bibfnamefont{C.}~\bibnamefont{{Heymans}}},
  \bibinfo{author}{\bibfnamefont{F.}~\bibnamefont{{Simpson}}},
  \bibnamefont{et~al.}, \bibinfo{journal}{\mnras}
  \textbf{\bibinfo{volume}{430}}, \bibinfo{pages}{2200} (\bibinfo{year}{2013}),
  \eprint{1212.3338}.

\bibitem[{\citenamefont{Smith et~al.}(2003)}]{Smith:2002dz}
\bibinfo{author}{\bibfnamefont{R.}~\bibnamefont{Smith}} \bibnamefont{et~al.}
  (\bibinfo{collaboration}{Virgo Consortium}), \bibinfo{journal}{\mnras}
  \textbf{\bibinfo{volume}{341}}, \bibinfo{pages}{1311} (\bibinfo{year}{2003}),
  \eprint{astro-ph/0207664}.

\bibitem[{\citenamefont{Takahashi et~al.}(2012)\citenamefont{Takahashi, Sato,
  Nishimichi, Taruya, and Oguri}}]{Takahashi:2012em}
\bibinfo{author}{\bibfnamefont{R.}~\bibnamefont{Takahashi}},
  \bibinfo{author}{\bibfnamefont{M.}~\bibnamefont{Sato}},
  \bibinfo{author}{\bibfnamefont{T.}~\bibnamefont{Nishimichi}},
  \bibinfo{author}{\bibfnamefont{A.}~\bibnamefont{Taruya}},
  \bibnamefont{et~al.}, \bibinfo{journal}{Astrophys.J.}
  \textbf{\bibinfo{volume}{761}}, \bibinfo{pages}{152} (\bibinfo{year}{2012}),
  \eprint{1208.2701}.

\bibitem[{\citenamefont{Bird et~al.}(2012)\citenamefont{Bird, Viel, and
  Haehnelt}}]{Bird:2011rb}
\bibinfo{author}{\bibfnamefont{S.}~\bibnamefont{Bird}},
  \bibinfo{author}{\bibfnamefont{M.}~\bibnamefont{Viel}}, \bibnamefont{and}
  \bibinfo{author}{\bibfnamefont{M.~G.} \bibnamefont{Haehnelt}},
  \bibinfo{journal}{Mon.Not.Roy.Astron.Soc.} \textbf{\bibinfo{volume}{420}},
  \bibinfo{pages}{2551} (\bibinfo{year}{2012}), \eprint{1109.4416}.

\bibitem[{\citenamefont{{Planck Collaboration}
  et~al.}(2013{\natexlab{e}})\citenamefont{{Planck Collaboration}, {Ade},
  {Aghanim}, {Armitage-Caplan}, {Arnaud}, {Ashdown}, {Atrio-Barandela},
  {Aumont}, {Baccigalupi}, {Banday} et~al.}}]{2013arXiv1303.5077P}
\bibinfo{author}{\bibnamefont{{Planck Collaboration}}},
  \bibinfo{author}{\bibfnamefont{P.~A.~R.} \bibnamefont{{Ade}}},
  \bibinfo{author}{\bibfnamefont{N.}~\bibnamefont{{Aghanim}}},
  \bibinfo{author}{\bibfnamefont{C.}~\bibnamefont{{Armitage-Caplan}}},
  \bibnamefont{et~al.}, \bibinfo{journal}{ArXiv e-prints}
  (\bibinfo{year}{2013}{\natexlab{e}}), \eprint{1303.5077}.

\bibitem[{\citenamefont{{van Engelen} et~al.}(2012)\citenamefont{{van Engelen},
  {Keisler}, {Zahn}, {Aird}, {Benson}, {Bleem}, {Carlstrom}, {Chang}, {Cho},
  {Crawford} et~al.}}]{2012ApJ...756..142V}
\bibinfo{author}{\bibfnamefont{A.}~\bibnamefont{{van Engelen}}},
  \bibinfo{author}{\bibfnamefont{R.}~\bibnamefont{{Keisler}}},
  \bibinfo{author}{\bibfnamefont{O.}~\bibnamefont{{Zahn}}},
  \bibinfo{author}{\bibfnamefont{K.~A.} \bibnamefont{{Aird}}},
  \bibnamefont{et~al.}, \bibinfo{journal}{\apj} \textbf{\bibinfo{volume}{756}},
  \bibinfo{eid}{142} (\bibinfo{year}{2012}), \eprint{1202.0546}.

\bibitem[{\citenamefont{Beutler et~al.}(2013)}]{Beutler:2013yhm}
\bibinfo{author}{\bibfnamefont{F.}~\bibnamefont{Beutler}} \bibnamefont{et~al.}
  (\bibinfo{collaboration}{BOSS Collaboration}) (\bibinfo{year}{2013}),
  \eprint{1312.4611}.

\bibitem[{\citenamefont{Beutler et~al.}(2014)}]{Beutler:2014yhv}
\bibinfo{author}{\bibfnamefont{F.}~\bibnamefont{Beutler}} \bibnamefont{et~al.}
  (\bibinfo{collaboration}{BOSS Collaboration}) (\bibinfo{year}{2014}),
  \eprint{1403.4599}.

\bibitem[{\citenamefont{{Lewis} et~al.}(2000)\citenamefont{{Lewis},
  {Challinor}, and {Lasenby}}}]{2000ApJ...538..473L}
\bibinfo{author}{\bibfnamefont{A.}~\bibnamefont{{Lewis}}},
  \bibinfo{author}{\bibfnamefont{A.}~\bibnamefont{{Challinor}}},
  \bibnamefont{and}
  \bibinfo{author}{\bibfnamefont{A.}~\bibnamefont{{Lasenby}}},
  \bibinfo{journal}{\apj} \textbf{\bibinfo{volume}{538}}, \bibinfo{pages}{473}
  (\bibinfo{year}{2000}), \eprint{arXiv:astro-ph/9911177}.

\bibitem[{\citenamefont{Taruya et~al.}(2012)\citenamefont{Taruya, Bernardeau,
  Nishimichi, and Codis}}]{Taruya:2012ut}
\bibinfo{author}{\bibfnamefont{A.}~\bibnamefont{Taruya}},
  \bibinfo{author}{\bibfnamefont{F.}~\bibnamefont{Bernardeau}},
  \bibinfo{author}{\bibfnamefont{T.}~\bibnamefont{Nishimichi}},
  \bibnamefont{and} \bibinfo{author}{\bibfnamefont{S.}~\bibnamefont{Codis}},
  \bibinfo{journal}{Phys.Rev.} \textbf{\bibinfo{volume}{D86}},
  \bibinfo{pages}{103528} (\bibinfo{year}{2012}), \eprint{1208.1191}.

\bibitem[{\citenamefont{Taruya et~al.}(2010)\citenamefont{Taruya, Nishimichi,
  and Saito}}]{Taruya:2010mx}
\bibinfo{author}{\bibfnamefont{A.}~\bibnamefont{Taruya}},
  \bibinfo{author}{\bibfnamefont{T.}~\bibnamefont{Nishimichi}},
  \bibnamefont{and} \bibinfo{author}{\bibfnamefont{S.}~\bibnamefont{Saito}},
  \bibinfo{journal}{Phys.Rev.} \textbf{\bibinfo{volume}{D82}},
  \bibinfo{pages}{063522} (\bibinfo{year}{2010}), \eprint{1006.0699}.

\bibitem[{\citenamefont{McDonald and Roy}(2009)}]{McDonald:2009dh}
\bibinfo{author}{\bibfnamefont{P.}~\bibnamefont{McDonald}} \bibnamefont{and}
  \bibinfo{author}{\bibfnamefont{A.}~\bibnamefont{Roy}},
  \bibinfo{journal}{JCAP} \textbf{\bibinfo{volume}{0908}}, \bibinfo{pages}{020}
  (\bibinfo{year}{2009}), \eprint{0902.0991}.

\bibitem[{\citenamefont{Rencher}(2003)}]{Rencher:2003}
\bibinfo{author}{\bibfnamefont{A.~C.} \bibnamefont{Rencher}},
  \emph{\bibinfo{title}{{Methods of Multivariate Analysis}}}
  (\bibinfo{publisher}{{John Wiley \& Sons}}, \bibinfo{year}{2003}),
  \bibinfo{edition}{2nd} ed.

\bibitem[{\citenamefont{Elgaroy and Lahav}(2005)}]{Elgaroy:2004}
\bibinfo{author}{\bibfnamefont{O.}~\bibnamefont{Elgaroy}} \bibnamefont{and}
  \bibinfo{author}{\bibfnamefont{O.}~\bibnamefont{Lahav}},
  \bibinfo{journal}{New J.Phys.} \textbf{\bibinfo{volume}{7}},
  \bibinfo{pages}{61} (\bibinfo{year}{2005}), \eprint{hep-ph/0412075}.

\bibitem[{\citenamefont{Lesgourgues and Pastor}(2006)}]{Lesgourgues:2006}
\bibinfo{author}{\bibfnamefont{J.}~\bibnamefont{Lesgourgues}} \bibnamefont{and}
  \bibinfo{author}{\bibfnamefont{S.}~\bibnamefont{Pastor}},
  \bibinfo{journal}{Phys.Rept.} \textbf{\bibinfo{volume}{429}},
  \bibinfo{pages}{307} (\bibinfo{year}{2006}), \eprint{astro-ph/0603494}.

\bibitem[{\citenamefont{Hannestad}(2010)}]{Hannestad:2010}
\bibinfo{author}{\bibfnamefont{S.}~\bibnamefont{Hannestad}},
  \bibinfo{journal}{Prog.Part.Nucl.Phys.} \textbf{\bibinfo{volume}{65}},
  \bibinfo{pages}{185} (\bibinfo{year}{2010}), \eprint{1007.0658}.

\bibitem[{\citenamefont{Hall and Challinor}(2012)}]{Hall:2012}
\bibinfo{author}{\bibfnamefont{A.~C.} \bibnamefont{Hall}} \bibnamefont{and}
  \bibinfo{author}{\bibfnamefont{A.}~\bibnamefont{Challinor}},
  \bibinfo{journal}{Mon.Not.Roy.Astron.Soc.} \textbf{\bibinfo{volume}{425}},
  \bibinfo{pages}{1170} (\bibinfo{year}{2012}), \eprint{1205.6172}.

\bibitem[{\citenamefont{Riess et~al.}(2011{\natexlab{b}})\citenamefont{Riess,
  Macri, Casertano, Lampeitl, Ferguson et~al.}}]{Riess:2011}
\bibinfo{author}{\bibfnamefont{A.~G.} \bibnamefont{Riess}},
  \bibinfo{author}{\bibfnamefont{L.}~\bibnamefont{Macri}},
  \bibinfo{author}{\bibfnamefont{S.}~\bibnamefont{Casertano}},
  \bibinfo{author}{\bibfnamefont{H.}~\bibnamefont{Lampeitl}},
  \bibnamefont{et~al.}, \bibinfo{journal}{Astrophys.J.}
  \textbf{\bibinfo{volume}{730}}, \bibinfo{pages}{119}
  (\bibinfo{year}{2011}{\natexlab{b}}), \eprint{1103.2976}.

\bibitem[{\citenamefont{Howlett et~al.}(2012)\citenamefont{Howlett, Lewis,
  Hall, and Challinor}}]{Howlett:2012}
\bibinfo{author}{\bibfnamefont{C.}~\bibnamefont{Howlett}},
  \bibinfo{author}{\bibfnamefont{A.}~\bibnamefont{Lewis}},
  \bibinfo{author}{\bibfnamefont{A.}~\bibnamefont{Hall}}, \bibnamefont{and}
  \bibinfo{author}{\bibfnamefont{A.}~\bibnamefont{Challinor}},
  \bibinfo{journal}{JCAP} \textbf{\bibinfo{volume}{1204}}, \bibinfo{pages}{027}
  (\bibinfo{year}{2012}), \eprint{1201.3654}.

\bibitem[{\citenamefont{Lewis and Challinor}(2002)}]{Lewis:2002}
\bibinfo{author}{\bibfnamefont{A.}~\bibnamefont{Lewis}} \bibnamefont{and}
  \bibinfo{author}{\bibfnamefont{A.}~\bibnamefont{Challinor}},
  \bibinfo{journal}{Phys.Rev.} \textbf{\bibinfo{volume}{D66}},
  \bibinfo{pages}{023531} (\bibinfo{year}{2002}), \eprint{astro-ph/0203507}.

\bibitem[{\citenamefont{Hu et~al.}(1998)\citenamefont{Hu, Eisenstein, and
  Tegmark}}]{Hu:1997}
\bibinfo{author}{\bibfnamefont{W.}~\bibnamefont{Hu}},
  \bibinfo{author}{\bibfnamefont{D.~J.} \bibnamefont{Eisenstein}},
  \bibnamefont{and} \bibinfo{author}{\bibfnamefont{M.}~\bibnamefont{Tegmark}},
  \bibinfo{journal}{Phys.Rev.Lett.} \textbf{\bibinfo{volume}{80}},
  \bibinfo{pages}{5255} (\bibinfo{year}{1998}), \eprint{astro-ph/9712057}.

\bibitem[{\citenamefont{Aguilar-Arevalo et~al.}(2001)}]{LSND}
\bibinfo{author}{\bibfnamefont{A.}~\bibnamefont{Aguilar-Arevalo}}
  \bibnamefont{et~al.} (\bibinfo{collaboration}{LSND Collaboration}),
  \bibinfo{journal}{Phys.Rev.} \textbf{\bibinfo{volume}{D64}},
  \bibinfo{pages}{112007} (\bibinfo{year}{2001}), \eprint{hep-ex/0104049}.

\bibitem[{\citenamefont{Armbruster et~al.}(2002)}]{KARMEN}
\bibinfo{author}{\bibfnamefont{B.}~\bibnamefont{Armbruster}}
  \bibnamefont{et~al.} (\bibinfo{collaboration}{KARMEN Collaboration}),
  \bibinfo{journal}{Phys.Rev.} \textbf{\bibinfo{volume}{D65}},
  \bibinfo{pages}{112001} (\bibinfo{year}{2002}), \eprint{hep-ex/0203021}.

\bibitem[{\citenamefont{Declais et~al.}(1995)\citenamefont{Declais, Favier,
  Metref, Pessard, Achkar et~al.}}]{Bugey}
\bibinfo{author}{\bibfnamefont{Y.}~\bibnamefont{Declais}},
  \bibinfo{author}{\bibfnamefont{J.}~\bibnamefont{Favier}},
  \bibinfo{author}{\bibfnamefont{A.}~\bibnamefont{Metref}},
  \bibinfo{author}{\bibfnamefont{H.}~\bibnamefont{Pessard}},
  \bibnamefont{et~al.}, \bibinfo{journal}{Nucl.Phys.}
  \textbf{\bibinfo{volume}{B434}}, \bibinfo{pages}{503} (\bibinfo{year}{1995}).

\bibitem[{\citenamefont{Aguilar-Arevalo et~al.}(2009)}]{MiniBooNE}
\bibinfo{author}{\bibfnamefont{A.}~\bibnamefont{Aguilar-Arevalo}}
  \bibnamefont{et~al.} (\bibinfo{collaboration}{MiniBooNE Collaboration}),
  \bibinfo{journal}{Phys.Rev.} \textbf{\bibinfo{volume}{D79}},
  \bibinfo{pages}{072002} (\bibinfo{year}{2009}), \eprint{0806.1449}.

\bibitem[{\citenamefont{Katori}(2014)}]{Katori:2014}
\bibinfo{author}{\bibfnamefont{T.}~\bibnamefont{Katori}}
  (\bibinfo{year}{2014}), \eprint{1404.6882}.

\bibitem[{\citenamefont{Mention et~al.}(2011)\citenamefont{Mention, Fechner,
  Lasserre, Mueller, Lhuillier et~al.}}]{Mention:2011}
\bibinfo{author}{\bibfnamefont{G.}~\bibnamefont{Mention}},
  \bibinfo{author}{\bibfnamefont{M.}~\bibnamefont{Fechner}},
  \bibinfo{author}{\bibfnamefont{T.}~\bibnamefont{Lasserre}},
  \bibinfo{author}{\bibfnamefont{T.}~\bibnamefont{Mueller}},
  \bibnamefont{et~al.}, \bibinfo{journal}{Phys.Rev.}
  \textbf{\bibinfo{volume}{D83}}, \bibinfo{pages}{073006}
  (\bibinfo{year}{2011}), \eprint{1101.2755}.

\bibitem[{\citenamefont{Giunti and Laveder}(2011)}]{Giunti:2010}
\bibinfo{author}{\bibfnamefont{C.}~\bibnamefont{Giunti}} \bibnamefont{and}
  \bibinfo{author}{\bibfnamefont{M.}~\bibnamefont{Laveder}},
  \bibinfo{journal}{Phys.Rev.} \textbf{\bibinfo{volume}{C83}},
  \bibinfo{pages}{065504} (\bibinfo{year}{2011}), \eprint{1006.3244}.

\bibitem[{\citenamefont{Archidiacono et~al.}(2013)\citenamefont{Archidiacono,
  Fornengo, Giunti, Hannestad, and Melchiorri}}]{Archidiacono:2013}
\bibinfo{author}{\bibfnamefont{M.}~\bibnamefont{Archidiacono}},
  \bibinfo{author}{\bibfnamefont{N.}~\bibnamefont{Fornengo}},
  \bibinfo{author}{\bibfnamefont{C.}~\bibnamefont{Giunti}},
  \bibinfo{author}{\bibfnamefont{S.}~\bibnamefont{Hannestad}},
  \bibnamefont{et~al.}, \bibinfo{journal}{Phys.Rev.}
  \textbf{\bibinfo{volume}{D87}}, \bibinfo{pages}{125034}
  (\bibinfo{year}{2013}), \eprint{1302.6720}.

\bibitem[{\citenamefont{Lesgourges et~al.}(2013)\citenamefont{Lesgourges,
  Mangano, Miele, and Pastor}}]{Les:nc}
\bibinfo{author}{\bibfnamefont{J.}~\bibnamefont{Lesgourges}},
  \bibinfo{author}{\bibfnamefont{G.}~\bibnamefont{Mangano}},
  \bibinfo{author}{\bibfnamefont{G.}~\bibnamefont{Miele}}, \bibnamefont{and}
  \bibinfo{author}{\bibfnamefont{S.}~\bibnamefont{Pastor}},
  \emph{\bibinfo{title}{{Neutrino Cosmology}}} (\bibinfo{publisher}{{Cambridge
  University Press}}, \bibinfo{address}{Cambridge}, \bibinfo{year}{2013}).

\bibitem[{\citenamefont{Dodelson and Widrow}(1994)}]{Dodelson:1993}
\bibinfo{author}{\bibfnamefont{S.}~\bibnamefont{Dodelson}} \bibnamefont{and}
  \bibinfo{author}{\bibfnamefont{L.~M.} \bibnamefont{Widrow}},
  \bibinfo{journal}{Phys.Rev.Lett.} \textbf{\bibinfo{volume}{72}},
  \bibinfo{pages}{17} (\bibinfo{year}{1994}), \eprint{hep-ph/9303287}.

\bibitem[{\citenamefont{Shi and Fuller}(1999)}]{Shi:1998}
\bibinfo{author}{\bibfnamefont{X.-D.} \bibnamefont{Shi}} \bibnamefont{and}
  \bibinfo{author}{\bibfnamefont{G.~M.} \bibnamefont{Fuller}},
  \bibinfo{journal}{Phys.Rev.Lett.} \textbf{\bibinfo{volume}{82}},
  \bibinfo{pages}{2832} (\bibinfo{year}{1999}), \eprint{astro-ph/9810076}.

\bibitem[{\citenamefont{{Gunn} and {Peterson}}(1965)}]{Gunn1965}
\bibinfo{author}{\bibfnamefont{J.~E.} \bibnamefont{{Gunn}}} \bibnamefont{and}
  \bibinfo{author}{\bibfnamefont{B.~A.} \bibnamefont{{Peterson}}},
  \bibinfo{journal}{\apj} \textbf{\bibinfo{volume}{142}}, \bibinfo{pages}{1633}
  (\bibinfo{year}{1965}),
  \adsurl{http://adsabs.harvard.edu/abs/1965ApJ...142.1633G}.

\bibitem[{\citenamefont{Becker et~al.}(2001)}]{Becker}
\bibinfo{author}{\bibfnamefont{R.~H.} \bibnamefont{Becker}}
  \bibnamefont{et~al.} (\bibinfo{collaboration}{SDSS Collaboration}),
  \bibinfo{journal}{Astron.J.} \textbf{\bibinfo{volume}{122}},
  \bibinfo{pages}{2850} (\bibinfo{year}{2001}), \eprint{astro-ph/0108097}.

\bibitem[{\citenamefont{Djorgovski et~al.}(2001)\citenamefont{Djorgovski,
  Castro, Stern, and Mahabal}}]{Djorgovski:2001}
\bibinfo{author}{\bibfnamefont{S.}~\bibnamefont{Djorgovski}},
  \bibinfo{author}{\bibfnamefont{S.}~\bibnamefont{Castro}},
  \bibinfo{author}{\bibfnamefont{D.}~\bibnamefont{Stern}}, \bibnamefont{and}
  \bibinfo{author}{\bibfnamefont{A.}~\bibnamefont{Mahabal}}
  (\bibinfo{year}{2001}), \eprint{astro-ph/0108069}.

\bibitem[{\citenamefont{Kaplinghat et~al.}(2003)\citenamefont{Kaplinghat, Chu,
  Haiman, Holder, Knox et~al.}}]{Kaplinghat}
\bibinfo{author}{\bibfnamefont{M.}~\bibnamefont{Kaplinghat}},
  \bibinfo{author}{\bibfnamefont{M.}~\bibnamefont{Chu}},
  \bibinfo{author}{\bibfnamefont{Z.}~\bibnamefont{Haiman}},
  \bibinfo{author}{\bibfnamefont{G.}~\bibnamefont{Holder}},
  \bibnamefont{et~al.}, \bibinfo{journal}{Astrophys.J.}
  \textbf{\bibinfo{volume}{583}}, \bibinfo{pages}{24} (\bibinfo{year}{2003}),
  \eprint{astro-ph/0207591}.

\bibitem[{\citenamefont{Kogut et~al.}(2003)}]{Kogut}
\bibinfo{author}{\bibfnamefont{A.}~\bibnamefont{Kogut}} \bibnamefont{et~al.}
  (\bibinfo{collaboration}{WMAP Collaboration}),
  \bibinfo{journal}{Astrophys.J.Suppl.} \textbf{\bibinfo{volume}{148}},
  \bibinfo{pages}{161} (\bibinfo{year}{2003}), \eprint{astro-ph/0302213}.

\bibitem[{\citenamefont{Elgaroy et~al.}(2002)\citenamefont{Elgaroy, Gramann,
  and Lahav}}]{Elgaroy:2001}
\bibinfo{author}{\bibfnamefont{O.}~\bibnamefont{Elgaroy}},
  \bibinfo{author}{\bibfnamefont{M.}~\bibnamefont{Gramann}}, \bibnamefont{and}
  \bibinfo{author}{\bibfnamefont{O.}~\bibnamefont{Lahav}},
  \bibinfo{journal}{Mon.Not.Roy.Astron.Soc.} \textbf{\bibinfo{volume}{333}},
  \bibinfo{pages}{93} (\bibinfo{year}{2002}), \eprint{astro-ph/0111208}.

\bibitem[{\citenamefont{Griffiths et~al.}(2001)\citenamefont{Griffiths, Silk,
  and Zaroubi}}]{Griffiths:2001}
\bibinfo{author}{\bibfnamefont{L.~M.} \bibnamefont{Griffiths}},
  \bibinfo{author}{\bibfnamefont{J.}~\bibnamefont{Silk}}, \bibnamefont{and}
  \bibinfo{author}{\bibfnamefont{S.}~\bibnamefont{Zaroubi}},
  \bibinfo{journal}{Mon.Not.Roy.Astron.Soc.} \textbf{\bibinfo{volume}{324}},
  \bibinfo{pages}{712} (\bibinfo{year}{2001}), \eprint{astro-ph/0010571}.

\bibitem[{\citenamefont{Hannestad et~al.}(2001)\citenamefont{Hannestad, Hansen,
  and Villante}}]{Hannestad:2001}
\bibinfo{author}{\bibfnamefont{S.}~\bibnamefont{Hannestad}},
  \bibinfo{author}{\bibfnamefont{S.}~\bibnamefont{Hansen}}, \bibnamefont{and}
  \bibinfo{author}{\bibfnamefont{F.}~\bibnamefont{Villante}},
  \bibinfo{journal}{Astropart.Phys.} \textbf{\bibinfo{volume}{16}},
  \bibinfo{pages}{137} (\bibinfo{year}{2001}), \eprint{astro-ph/0012009}.

\bibitem[{\citenamefont{Barriga et~al.}(2001)\citenamefont{Barriga, Gaztanaga,
  Santos, and Sarkar}}]{Barriga:2001}
\bibinfo{author}{\bibfnamefont{J.}~\bibnamefont{Barriga}},
  \bibinfo{author}{\bibfnamefont{E.}~\bibnamefont{Gaztanaga}},
  \bibinfo{author}{\bibfnamefont{M.}~\bibnamefont{Santos}}, \bibnamefont{and}
  \bibinfo{author}{\bibfnamefont{S.}~\bibnamefont{Sarkar}},
  \bibinfo{journal}{Mon.Not.Roy.Astron.Soc.} \textbf{\bibinfo{volume}{324}},
  \bibinfo{pages}{977} (\bibinfo{year}{2001}), \eprint{astro-ph/0011398}.

\bibitem[{\citenamefont{Gramann and Hutsi}(2001)}]{Gramann:2001}
\bibinfo{author}{\bibfnamefont{M.}~\bibnamefont{Gramann}} \bibnamefont{and}
  \bibinfo{author}{\bibfnamefont{G.}~\bibnamefont{Hutsi}},
  \bibinfo{journal}{Mon.Not.Roy.Astron.Soc.} \textbf{\bibinfo{volume}{327}},
  \bibinfo{pages}{538} (\bibinfo{year}{2001}), \eprint{astro-ph/0102466}.

\bibitem[{\citenamefont{Kaiser et~al.}(2013)\citenamefont{Kaiser, Mazenc, and
  Sfakianakis}}]{Kaiser:2012}
\bibinfo{author}{\bibfnamefont{D.~I.} \bibnamefont{Kaiser}},
  \bibinfo{author}{\bibfnamefont{E.~A.} \bibnamefont{Mazenc}},
  \bibnamefont{and} \bibinfo{author}{\bibfnamefont{E.~I.}
  \bibnamefont{Sfakianakis}}, \bibinfo{journal}{Phys.Rev.}
  \textbf{\bibinfo{volume}{D87}}, \bibinfo{pages}{064004}
  (\bibinfo{year}{2013}), \eprint{1210.7487}.

\bibitem[{\citenamefont{Benetti et~al.}(2013)\citenamefont{Benetti, Pandolfi,
  Lattanzi, Martinelli, and Melchiorri}}]{Benetti:2013a}
\bibinfo{author}{\bibfnamefont{M.}~\bibnamefont{Benetti}},
  \bibinfo{author}{\bibfnamefont{S.}~\bibnamefont{Pandolfi}},
  \bibinfo{author}{\bibfnamefont{M.}~\bibnamefont{Lattanzi}},
  \bibinfo{author}{\bibfnamefont{M.}~\bibnamefont{Martinelli}},
  \bibnamefont{et~al.}, \bibinfo{journal}{Phys.Rev.}
  \textbf{\bibinfo{volume}{D87}}, \bibinfo{pages}{023519}
  (\bibinfo{year}{2013}), \eprint{1210.3562}.

\bibitem[{\citenamefont{Benetti}(2013)}]{Benetti:2013b}
\bibinfo{author}{\bibfnamefont{M.}~\bibnamefont{Benetti}},
  \bibinfo{journal}{Phys.Rev.} \textbf{\bibinfo{volume}{D88}},
  \bibinfo{pages}{087302} (\bibinfo{year}{2013}), \eprint{1308.6406}.

\bibitem[{\citenamefont{Hazra et~al.}(2013)\citenamefont{Hazra, Shafieloo, and
  Smoot}}]{Hazra:2013}
\bibinfo{author}{\bibfnamefont{D.~K.} \bibnamefont{Hazra}},
  \bibinfo{author}{\bibfnamefont{A.}~\bibnamefont{Shafieloo}},
  \bibnamefont{and} \bibinfo{author}{\bibfnamefont{G.~F.} \bibnamefont{Smoot}},
  \bibinfo{journal}{JCAP} \textbf{\bibinfo{volume}{1312}}, \bibinfo{pages}{035}
  (\bibinfo{year}{2013}), \eprint{1310.3038}.

\bibitem[{\citenamefont{Starobinsky}(1992)}]{Starobinsky:1992}
\bibinfo{author}{\bibfnamefont{A.~A.} \bibnamefont{Starobinsky}},
  \bibinfo{journal}{JETP Lett.} \textbf{\bibinfo{volume}{55}},
  \bibinfo{pages}{489} (\bibinfo{year}{1992}).

\bibitem[{\citenamefont{Calabrese et~al.}(2008)\citenamefont{Calabrese, Slosar,
  Melchiorri, Smoot, and Zahn}}]{Calabrese:2008}
\bibinfo{author}{\bibfnamefont{E.}~\bibnamefont{Calabrese}},
  \bibinfo{author}{\bibfnamefont{A.}~\bibnamefont{Slosar}},
  \bibinfo{author}{\bibfnamefont{A.}~\bibnamefont{Melchiorri}},
  \bibinfo{author}{\bibfnamefont{G.~F.} \bibnamefont{Smoot}},
  \bibnamefont{et~al.}, \bibinfo{journal}{Phys.Rev.}
  \textbf{\bibinfo{volume}{D77}}, \bibinfo{pages}{123531}
  (\bibinfo{year}{2008}), \eprint{0803.2309}.

\bibitem[{\citenamefont{Jeffreys}(1973)}]{Jeffreys:1973}
\bibinfo{author}{\bibfnamefont{S.~H.} \bibnamefont{Jeffreys}},
  \emph{\bibinfo{title}{Scientific Inference}} (\bibinfo{publisher}{Cambridge
  University Press}, \bibinfo{address}{Cambridge, UK}, \bibinfo{year}{1973}),
  \bibinfo{edition}{3rd} ed.

\bibitem[{\citenamefont{Verde et~al.}(2013)\citenamefont{Verde, Feeney,
  Mortlock, and Peiris}}]{Verde:2013}
\bibinfo{author}{\bibfnamefont{L.}~\bibnamefont{Verde}},
  \bibinfo{author}{\bibfnamefont{S.~M.} \bibnamefont{Feeney}},
  \bibinfo{author}{\bibfnamefont{D.~J.} \bibnamefont{Mortlock}},
  \bibnamefont{and} \bibinfo{author}{\bibfnamefont{H.~V.}
  \bibnamefont{Peiris}}, \bibinfo{journal}{JCAP}
  \textbf{\bibinfo{volume}{1309}}, \bibinfo{pages}{013} (\bibinfo{year}{2013}),
  \eprint{1307.2904}.

\bibitem[{\citenamefont{Jeffreys}(1961)}]{Jeffreys:1961}
\bibinfo{author}{\bibfnamefont{S.~H.} \bibnamefont{Jeffreys}},
  \emph{\bibinfo{title}{The Theory of Probability}} (\bibinfo{publisher}{Oxford
  University Press}, \bibinfo{address}{New York, NY, USA},
  \bibinfo{year}{1961}).

\bibitem[{\citenamefont{Leistedt et~al.}(2014)\citenamefont{Leistedt, Peiris,
  and Verde}}]{Boris}
\bibinfo{author}{\bibfnamefont{B.}~\bibnamefont{Leistedt}},
  \bibinfo{author}{\bibfnamefont{H.~V.} \bibnamefont{Peiris}},
  \bibnamefont{and} \bibinfo{author}{\bibfnamefont{L.}~\bibnamefont{Verde}}
  (\bibinfo{year}{2014}), \eprint{1404.5950}.

\bibitem[{\citenamefont{Capozzi et~al.}(2013)\citenamefont{Capozzi, Fogli,
  Lisi, Marrone, Montanino et~al.}}]{Capozzi:2013csa}
\bibinfo{author}{\bibfnamefont{F.}~\bibnamefont{Capozzi}},
  \bibinfo{author}{\bibfnamefont{G.}~\bibnamefont{Fogli}},
  \bibinfo{author}{\bibfnamefont{E.}~\bibnamefont{Lisi}},
  \bibinfo{author}{\bibfnamefont{A.}~\bibnamefont{Marrone}},
  \bibnamefont{et~al.} (\bibinfo{year}{2013}), \eprint{1312.2878}.

\bibitem[{\citenamefont{Weinberg et~al.}(2013)\citenamefont{Weinberg, Bard,
  Dawson, Dore, Frieman et~al.}}]{Weinberg:2013raj}
\bibinfo{author}{\bibfnamefont{D.}~\bibnamefont{Weinberg}},
  \bibinfo{author}{\bibfnamefont{D.}~\bibnamefont{Bard}},
  \bibinfo{author}{\bibfnamefont{K.}~\bibnamefont{Dawson}},
  \bibinfo{author}{\bibfnamefont{O.}~\bibnamefont{Dore}}, \bibnamefont{et~al.}
  (\bibinfo{year}{2013}), \eprint{1309.5380}.

\bibitem[{\citenamefont{Albrecht et~al.}(2006)\citenamefont{Albrecht,
  Bernstein, Cahn, Freedman, Hewitt et~al.}}]{Albrecht:2006um}
\bibinfo{author}{\bibfnamefont{A.}~\bibnamefont{Albrecht}},
  \bibinfo{author}{\bibfnamefont{G.}~\bibnamefont{Bernstein}},
  \bibinfo{author}{\bibfnamefont{R.}~\bibnamefont{Cahn}},
  \bibinfo{author}{\bibfnamefont{W.~L.} \bibnamefont{Freedman}},
  \bibnamefont{et~al.} (\bibinfo{year}{2006}), \eprint{astro-ph/0609591}.

\bibitem[{\citenamefont{{Lewis} and {Bridle}}(2002)}]{2002PhRvD..66j3511L}
\bibinfo{author}{\bibfnamefont{A.}~\bibnamefont{{Lewis}}} \bibnamefont{and}
  \bibinfo{author}{\bibfnamefont{S.}~\bibnamefont{{Bridle}}},
  \bibinfo{journal}{\prd} \textbf{\bibinfo{volume}{66}}, \bibinfo{eid}{103511}
  (\bibinfo{year}{2002}), \eprint{arXiv:astro-ph/0205436}.

\end{thebibliography}

\end{document}